\documentclass[12pt,preprint]{aastex}

\usepackage{graphicx}
\usepackage{epstopdf}
\usepackage{amsmath}

\def\gs{\mathrel{\raise0.35ex\hbox{$\scriptstyle >$}\kern-0.6em
\lower0.40ex\hbox{{$\scriptstyle \sim$}}}}
\def\ls{\mathrel{\raise0.35ex\hbox{$\scriptstyle <$}\kern-0.6em
\lower0.40ex\hbox{{$\scriptstyle \sim$}}}}

\newcommand{\um}{\,$\mu$m}

\newcommand{\lsun}{\,$\rm{L}_{\odot}$}
\newcommand{\msun}{\,$\rm{M}_{\odot}$}

\slugcomment{DRAFT \today}

\shorttitle{Mid-IR absorption features at $z$\,$\sim$\,2}
\shortauthors{Sajina et al.}

\begin{document}

\title{Detections of water ice, hydrocarbons, and 3.3\um\ PAH in $z$\,$\sim$\,2 ULIRGs}

\author{Anna Sajina$^1$,  Henrik Spoon$^2$, Lin Yan$^3$, Masatoshi Imanishi$^4$,Dario Fadda$^5$,Moshe Elitzur$^6$}

\affil{$^1$ Physics \& Astronomy Department, Haverford College, Haverford, PA, 19041, USA}
\affil{$^2$ Astronomy Department, Cornell University, Ithaca, NY 14853, USA}
\affil{$^3$ IPAC, California Institute of Technology, Pasadena, CA 91125, USA}
\affil{$4$ National Astronomical Observatory, 2-21-1, Osawa, Mitaka, Tokyo 181-8588, Japan}
\affil{$^5$ NASA {\sl Herschel} Space Telescope Center, California Institute of Technology, Pasadena, CA 91125, USA}
\affil{$^6$ Department of Physics and Astronomy, University of Kentucky,  Lexington, KY 40506, USA}
\begin{abstract}

We present the first detections of the 3\um\ water ice and 3.4\um\ amorphous hydrocarbon (HAC) absorption features in $z$\,$\sim$\,2 ULIRGs. These are based on deep rest-frame 2\,--\,8\um\ {\sl Spitzer} IRS spectra of 11 sources selected for their appreciable silicate absorption. The HAC-to-silicate ratio for our $z$\,$\sim$\,2 sources is typically higher by a factor of 2\,--\,5 than that observed in the Milky Way.  This HAC `excess'  suggests compact nuclei with steep temperature gradients as opposed to predominantly host obscuration. Beside the above molecular absorption features, we detect the 3.3\um\ PAH emission feature in one of our sources with three more individual spectra showing evidence for it. Stacking analysis suggests that water ice, hydrocarbons, and PAH are likely present in the bulk of this sample even when not individually detected.  
The most unexpected result of our study is the lack of clear detections of the 4.67\um\ CO gas absorption feature. Only three of the sources show tentative signs of this feature and at significantly lower levels than has been observed in local ULIRGs. Overall, we find that the closest local analogs to our sources, in terms of 3\,--\,4\um\ color, HAC-to-silicate and ice-to-silicate ratios, as well as low PAH equivalent widths are sources dominated by deeply obscured nuclei. Such sources form only a small fraction of ULIRGs locally and are commonly believed to be dominated by buried AGN.  Our sample suggests that, in absolute number, such buried AGN are at least an order of magnitude more common at $z$\,$\sim$\,2 than today. The presence of PAH suggests that significant levels of star-formation are present even if the obscured AGN typically dominate the power budget.

\end{abstract}

\keywords{galaxies:high-redshift,galaxies:ISM, infrared:galaxies,quasars:absorption lines}

\section{Introduction}

$\indent$One of the most exciting recent discoveries in astronomy is the strong evolution and co-evolution of both the star-formation rate density, and black hole number density from $z$\,$\sim$\,1\,--\,3 to today \citep[see e.g.][]{wall05,hop06,phop08}.  Particularly interesting are composite or transition objects where quasars co-exist with powerful starbursts \citep[see e.g.][]{cs01,coppin08,walter09}.  Such objects are likely to be much more common at $z$\,$\sim$\,2, the peak epoch of both the quasar number density and the star-formation rate density. In particular, recent {\sl Spitzer} Space Telescope \citep{werner04} results suggest that the ULIRG number density was more than 200-fold greater at $z$\,$\sim$\,2 than it is today \citep[for a review see][]{soifer08}. This strong evolution of the luminosity function means that many more extremely luminous sources ($L_{\rm{IR}}$\,$\sim$\,$10^{13}$\lsun) have been discovered than are observed locally. By necessity therefore these extreme sources are typically compared with local ULIRGs which are in fact much less luminous (typically $L_{\rm{IR}}$\,$\sim$\,$10^{12}$\,--\,$10^{12.4}$\lsun). Some of these $z$\,$\sim$\,2 sources derive their luminosities from powerful obscured quasars, but others appear to be scaled-up versions of the local largely starburst-dominated ULIRGs \citep[e.g.][]{sajina07,sajina08,polletta08,hc09}. Due to the dearth of such sources nearby the physical conditions such as the dust composition at these highest luminosities is poorly understood. 

One approach is to determine whether or not the detectability and relative strength of absorption and emission features arising from different dust components are consistent with those of local ULIRGs in order to test how similar are their dust properties.  The mid-IR regime is well suited to such a study as it contains a wealth of molecular absorption features. By far the strongest is the 9.7\um\ silicate absorption feature (Si-O stretching mode). This is thought to sample the large silicate grains, which are frequently coated in various ices including water. These ices can be seen in the H-O-H bending mode (3.05\um) and the O-H stretching mode (6.15\um).  In more extreme environments such as observed toward the Galactic Center, the degree of hydrogenation is such that prominent hydrocarbon (HAC) features are observed \citep[e.g.][]{pendleton94,whittet03}.  The strongest of these is the 3.4\um\ C-H stretching mode, but also includes the 6.85 and 7.25\um\ C-H deformation modes.  Studies of these features in external galaxies and in particular ULIRGs are still quite challenging. The first such study was that of \citet{spoon02} who, out of a sample of 100 sources with {\sl ISO}-SWS spectra, detected the 6\um\ water ice absorption in 18 galaxies (including most of the ULIRGs).  Indeed ULIRGs show the deepest silicate absorption features \citep[e.g.][]{hao07,spoon07,sirocky08}, and hence are the best candidates to look for the other mid-IR molecular absorption features.  The Subaru study of the 3\,--\,4\um\ spectra of 37 ULIRGs \citep{imanishi_ulirgs} supplemented by the $\sim$\,5\,--\,30\um\ {\sl Spitzer} Infrared Spectrograph \citep[IRS;][]{houck04} study of local ULIRGs  \citep{imanishi_irs}, and the VLT study of the 3\,--\,4\um\ spectra of 11 more ULIRGs by \citet{risaliti06} all show molecular absorption features.  Over the 3\,--\,5\um\ regime, the VLT ISAAC spectra of 5 ULIRGs were presented in \citet{sani08}. Most recently {\sl AKARI} spectra of 45 local ULIRGs provide the first continuous 2.5\,--\,5\um\  spectra \citep{imanishi08} of a large number of ULIRGs and serve as an excellent local basis for comparison with our high-$z$ spectra. Lastly, CO gas has a prominent 4.67\um\ absorption feature, which has been detected in several deeply obscured local ULIRGs \citep{spoon_f00183,spoon_co,imanishi08,sani08}. An archival study of {\sl ISO} spectra of Seyfert galaxies showed no sign of this feature, presumably due to different obscuration geometries, i.e. dust torus vs. buried nucleus \citep{lutz04}. All these absorption features arise in slightly different conditions and hence sample different obscuring layers \citep{spoon_f00183}.  However, how these features arise and their relative strengths are still not well understood even locally. Only a few dust models that address the relative strength of the different absorption features have been published \citep[e.g.][]{oss94,zinoveva05}, which along with laboratory ice spectra \citep[e.g.][]{icelab} help in interpreting the results. Most of what we know about interstellar ices comes from observations of embedded protostars \citep[for a review see][]{gibb04} which in may also be analogs to our most highly obscured $z$\,$\sim$\,2 sources than local ULIRGs.
 
In this paper, we present unprecedented signal-to-noise rest-frame $\sim$\,2\,--\,8\um\ spectra of $z$\,$\sim$\,2 ULIRGs. A spectrum of a $z$\,$\sim$\,2.56 galaxy with similar signal-to-noise is presented in \citet{rigby08}, but starting at restframe $\sim$\,5.5\um. Spectra extending to the rest-frame $\sim$\,2\um\ have several advantages which will shed light on the physical conditions of these distant systems: 1) the continuum arises from hot dust near the power source, 2) this regime includes the 3.3\um\ and 6.2\um\ PAH features -- both isolated and less severely affected by the mid-IR continuum model than the longer wavelength PAH features, 3) dust absorption features including water ice at 3\um\ and 6\um, several hydrocarbon (HAC) features, and 4) the warm CO gas 4.67\um\  absorption feature. In this paper we concentrate on the various absorption and emission features in these  spectra and compare our results with the available local ULIRG data. Throughout this paper we assume a flat Universe with $\Omega_{\rm{M}}$\,=\,0.27, $\Omega_{\Lambda}$\,=\,0.73, and $H_0$\,=\,71\,km\,s$^{-1}$\,Mpc$^{-1}$ \citep{spergel03}.

\section{Sample}
\subsection{Background}
$\indent$Our group has been engaged in a multifaceted study of the  properties of high-$z$ ULIRGs, including low-resolution IRS spectra, {\sl HST} imaging, $X$-ray imaging, and Keck spectroscopy. We have $\sim$\,200 IRS spectra of sources with $F_{24}$\,$>$\,0.8\,mJy, selected in the {\sl Spitzer} First Look Survey field\footnote{http://ssc.spitzer.caltech.edu/fls/}, of which $\sim$\,70 sources are at 1.5\,$<$\,$z$\,$<$\,2.5, and likely $L_{\rm{IR}}$\,$\sim$\,$10^{13}$\lsun\ \citep{yan07}.  The analysis of the first 50 sources (GO1 program \#3768) of this sample has already been completed \citep{yan07,sajina07}. The remaining 150 sources (GO2 program \#20629) give a largely uniform, flux-limited sample (Dasyra et al. in prep.). Both the G01 and G02 samples have a limit of $F_{24}$\,$>$\,0.8\,mJy, but our G01 sources  have two additional selection criteria: $\nu F_{\nu}$(24\um)/$\nu F_{\nu}$(0.64\um)\,$>$\,1.0 and $\nu F_{\nu}$(24)/$\nu F_{\nu}$(8)\,$>$\,0.5. The later selection cut means that the G01 sources tend to be fainter at 8\um\ (rest-frame 2\um) than the G02 sources. 

\subsection{Target selection}
We adopt three selection criteria for our follow-up program: 1) redshift, 2) brightness, and 3) silicate feature depth. We adopt the redshift range 1.6\,$<$\,$z$\,$<$\,2.5.  These boundaries are determined by practical reasons: beyond $z$\,$>$\,2.5, IRS spectra poorly sample the 9.7\um\ silicate feature, while sources below $z$\,$\sim$\,1.6 require an observationally-expensive additional IRS module  (SL2).  For the brightness, we simply require an 8\um\ detection ($\gs$\,60\,$\mu$Jy) or else the SL spectra would be too difficult to obtain.  The last criterion was most difficult to define. As this is a largely exploratory program, we did not want to restrict ourselves to only the most extremely obscured sources \citep[see e.g. category 3;][]{spoon07}, but also needed the sources to have relatively strong silicate absorption features since, as expected, only sources with silicate absorption have been observed to have ice absorption features \citep{imanishi_irs}.  We selected, $\tau_{9.7}$\,$>$\,2.0 (based on the Galactic Center extinction curve used in Sajina et al. 2007a) as a rather generous lower limit. In terms of the observed 9.7\um\ depth (i.e. $\ln(F_{\rm{cont}}/F_{\rm{obs}})$  as is more commonly used, this translates to $\tau_{Si}$\,$\equiv$\,$\tau_{9.7}$\,$>$\,1.45. For the sake of comparison with previous work, throughout this paper we use observed optical depths (see also \S\,\ref{sec_optdepth}). Out of 52 sources in the GO1 sample,  6 sources satisfy the above conditions.  Of the G02 sample, 5 sources satisfy the conditions.  However, as we used very preliminary data reduction of this sample, when doing the initial sample selection, one of the sources that meets the criteria (MIPS509) was not included, and another (MIPS22633) was observed although its silicate feature depth is too low compared with the rest ($\tau_{9.7}$\,$\sim$\,1.1).  In total, our sample consists of 11 sources. 

\subsection{Redshift accuracy}
Only 3/11  of our sources have optical/near-IR redshifts (see Yan et al. 2007 and Sajina et al. 2008). The rest are typically based on the 10\um\ silicate absorption feature and hence are believed to have an uncertainty, $\Delta z$\,$\sim$\,0.2 (see Yan et al. 2007). Hence, for the sources without accurate redshifts, we look for example for absorption features within the range $z_{\rm{IRS}}$\,$\pm$\,$\Delta z$. This constraint makes the reliability of single marginally-detected features doubtful. In cases of unambiguous and/or multiple absorption feature detections, these can be used to improve our redshift estimates. As the rest-frame peak of the molecular absorption features can also vary -- for example the 3\um\ ice feature can peak anywhere in the range $\sim$\,3.05\,--\, 3.1\um\ \citep{chiar02,zinoveva05}, an uncertainty of up to $\Delta z$\,$\sim$\,0.05 remains. 
 
\section{IRS observations}

In terms of both the molecular absorption features and hot dust continuum, we are interested in obtaining deeper spectra in the rest-frame $\sim$\,2\,--\,8\um\ spectral range. At the redshifts of our targets this means using the {\sl Spitzer} IRS's SL1 and LL2 modules. 
One of our targets, MIPS22432 is very bright ($F_8$\,=\,0.5\,mJy), and was already observed in SL1 as part of our GO2 program. Its signal-to-noise is sufficient for our purposes (see below) and hence does not need to be re-observed in SL1.  Both MIPS22432 and MIPS15880 have sufficiently high signal-to-noise and do not need new LL2 observations. Lastly, MIPS8392, was already observed in SL1 as part of the G01 program \#15 (P.I. Houck) and hence its SL1 observation was removed from our observations list.  Unfortunately, the depth of the previous observation is 3\,$\times$ lower than we needed here, but this is still considered a duplicate observation (the limit is 4\,$\times$). The depth of the earlier observation is insufficient for our purposes hence it can serve only to constrain the continuum levels, but is not useful in looking for absorption features. 

Table~\ref{table_obs} gives a summary of the observations covering three {\sl Spitzer} observing cycles (G01,G02, and G04).  The full GO1 sample and observations are discussed in detail in \citet{yan07}. The full GO2 sample and observations are discussed in \citet{dasyra09}. The GO4 observations (PID\#40793) are presented here for the first time.  Combining all these programs, each source has typically $\sim$\,2.5\,hrs on source of IRS SL1 and LL2 observations, although brighter sources have less exposure time as can be seen in Table~\ref{table_obs}.   The SL1 observations were mostly in mapping mode in order to improve the signal-to-noise ratio in these very faint sources.  For our sources, we adopted the IRAC and/or VLA positions which are the most accurate available and sufficiently accurate for the SL1 module which has the narrowest slit width (3.6\arcsec\ compared with 10.5\arcsec\ for LL2).  

\subsection{Data reduction}

We use IrsLow, a data reduction code for IRS low resolution observations custom-written by Dario Fadda (see Fadda et al. 2010 in prep. for further details). The data reduction steps include bad pixel removal, background subtraction, co-adding different positions (such as the 2 nod positions in stare mode, or the 4 positions we typically used in map mode), and finally extracting the spectra using optimal extraction. Bad pixels are determined as outliers across the available frames. For the SL data for example, fainter sources have 40 frames per source, but the bright sources have only 6 frames per source. Therefore this bad pixel flagging is less efficient for our bright sources with very few individual frames leading to somewhat noisier spectra than the nominal difference in exposure time. The code is iterative in that after the first background subtraction, any sources present are masked, and a second background subtraction is done. In addition, we test the reliability of any features we see, by displaying all points that go into the spectral extraction and manually clipping any obvious outliers (on top of the bad pixel flagging of the code) that might bias our results. A few spurious spikes are removed in this way, but  this is mostly a concern for narrow features of interest such as the 3.3\um\ PAH feature. In all cases, where we believe that this feature is or might be present we have used this option of the code to check for any possible outliers. For the broad water ice or HAC absorption features, the quality of the background subtraction is more of a concern than any remaining outlier pixels. Below we describe our background subtraction quality test.

As a double check on our per pixel errors (and the background subtraction) we also extract sky spectra from a position in the data image which does not include a source. The same PSF is used for the off-source spectra extraction as the source spectra. These sky spectra allow us to ascertain that the mean background level is zero and that no residual gradients are present.  The mean rms values from these sky spectra are found to be comparable with the typical per pixel errors of the extracted spectra (but $\sim$\,30\% worse in the case of the two SL stare mode observations due to the less efficient bad pixel flagging).  For the SL1 spectra the rms is typically $\sim$\,20\,--\,50$\mu$Jy (with the lower values typical of the fainter sources and higher values for the brighter sources) with continuum SNRs of $\sim$\,4--8. For the LL2 spectra the rms is typically $\sim$\,60\,$\mu$Jy with continuum SNRs ranging from  $\sim$\,3 for MIPS22530 to $\sim$\,25 for MIPS22303, with typical values of $\sim$\,10.

We tested IrsLow's spectral extraction, by running SPICE after the background subtraction and obtained fully consistent results.  We find that our old reduction of the LL2 data for the GO1 sources \citep{yan07} is frequently lower than the new data by up to 30\% (somewhat worse for MIPS464). However, when re-reduced the same way as the new data (see above) the two sets are consistent. We attribute the difference to improved background subtraction and calibration.  For consistency we have re-reduced all available data (including LL1, LL2 and SL1) before co-adding into the final spectra. 

\section{Results}
 \subsection{Individual spectra}
Figure\,\ref{fig_specs} shows the eleven spectra of our sources derived by co-adding all available data (see Table~\ref{table_obs}).  For the LL2 region ($\lambda_{\rm{obs}}$\,$\sim$\,14\,--\,21\um), the spectra represent the weighted mean of the old and new data. The panels below show closer views of the spectra around the principal water ice absorption features at 3\um\ and 6\um. We detect the continuum down to $\sim$\,2.5\um\ allowing us to look for the various molecular absorption features expected in this regime. Indeed, we see a number of interesting features in the spectra, the vast majority of which can be identified with the known molecular features including water ice, PAH, HAC, and CO. Figure\,\ref{fig_specs} also shows the SL1 and LL2 sky spectra for these sources (i.e. spectra taken off-source, but with the same PSF). As expected, these sky spectra have a median of essentially zero, but do show some `bumps' and `dips' which in addition to the background noise, likely also include some low-level rogue pixels that are not sufficiently strong outliers to have been removed. Such spurious features, however, could influence our source spectra as well. Essentially there are three considerations for believing the reality of a given feature: 1) whether or not it is stronger than the amplitude of the features in the sky spectra, 2) whether or not it has consistent redshift, and 3) whether or not it has roughly the expected shape (i.e. width). We address these issues in \S\,\ref{sec_optdepth}. Before hand however, we examine briefly the hot dust continuum shape of the sources. 

\begin{figure*}[t]
\centering \leavevmode
\includegraphics[height=6cm]{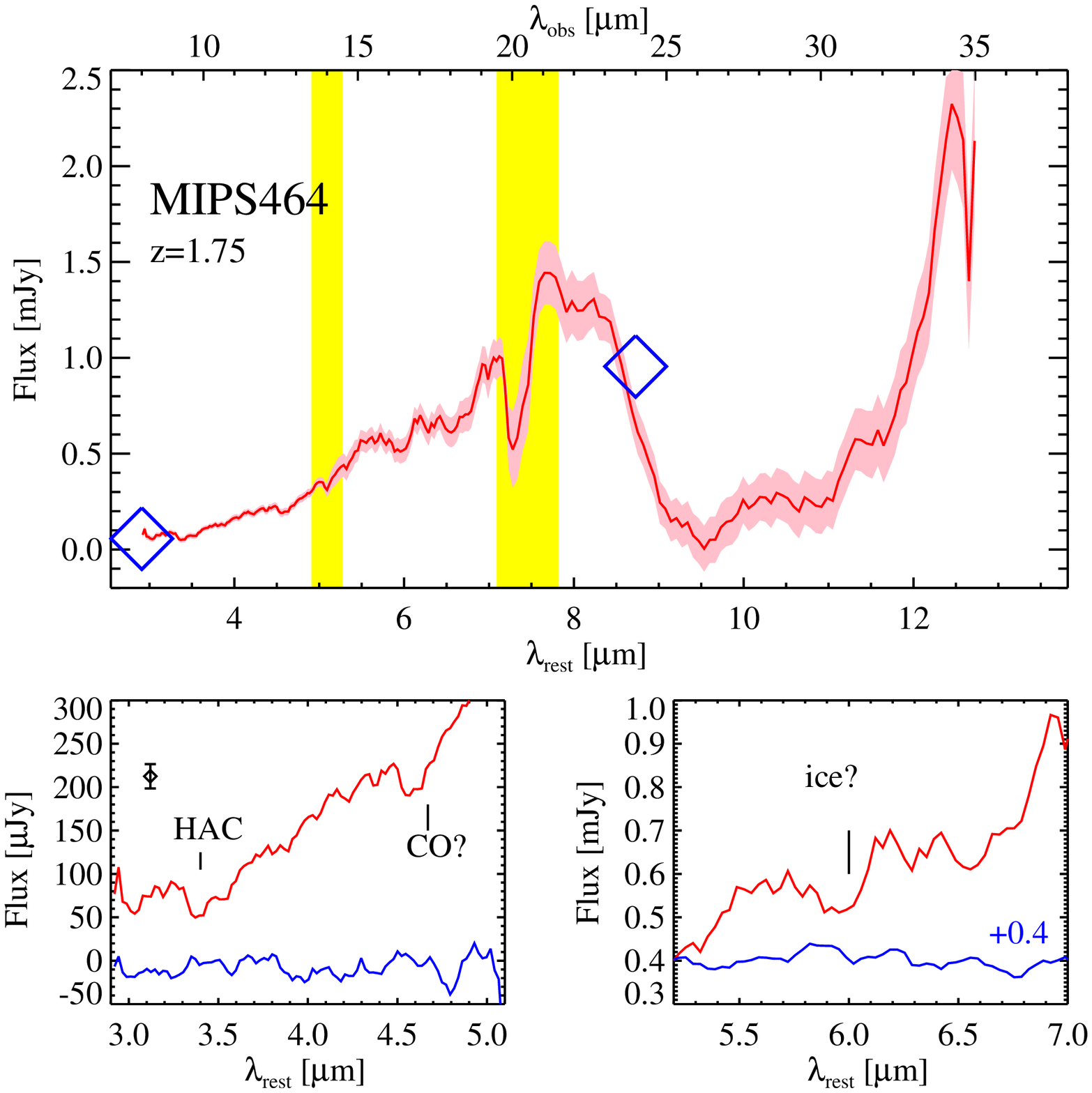}
\includegraphics[height=6cm]{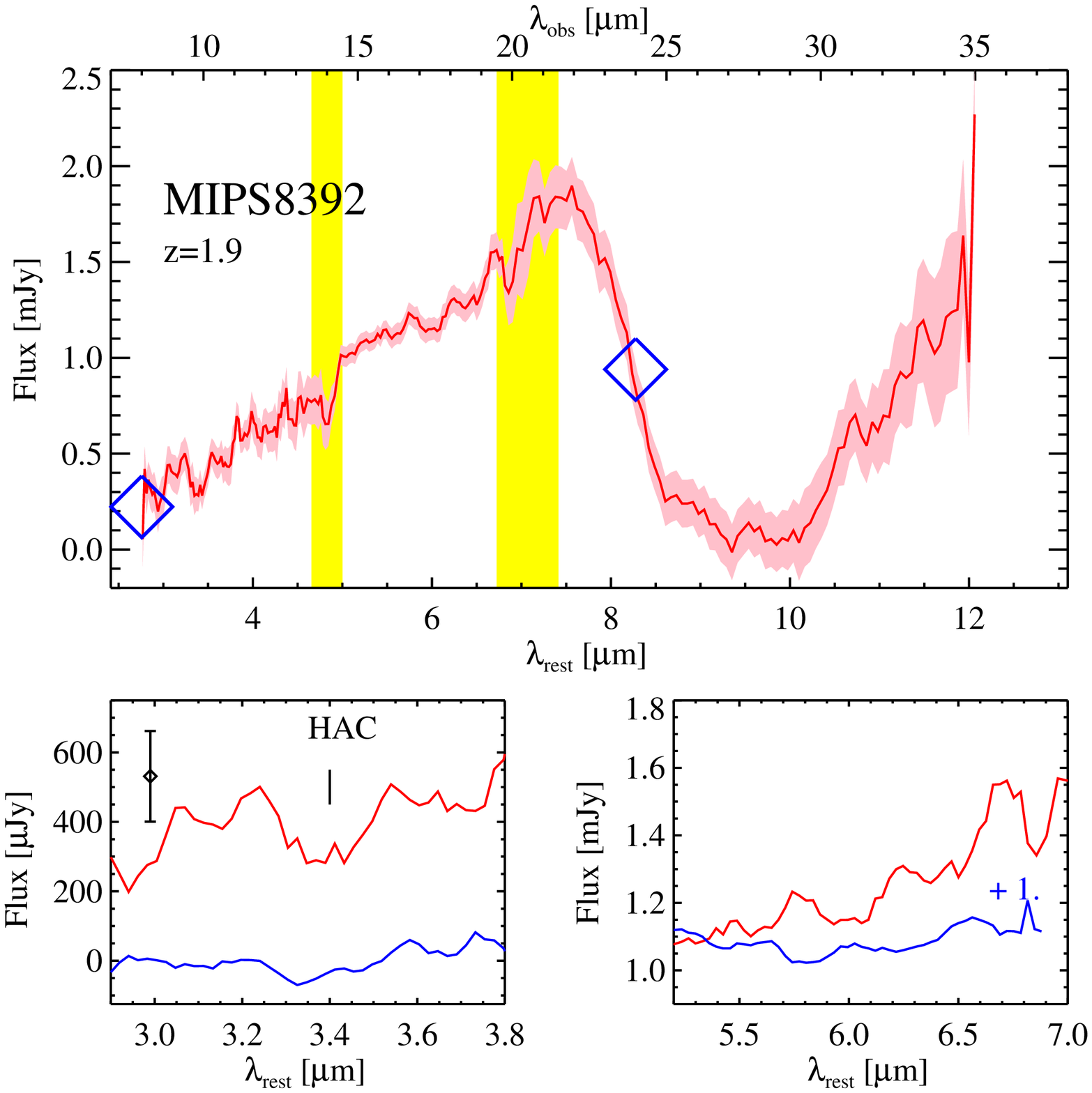}
\includegraphics[height=6cm]{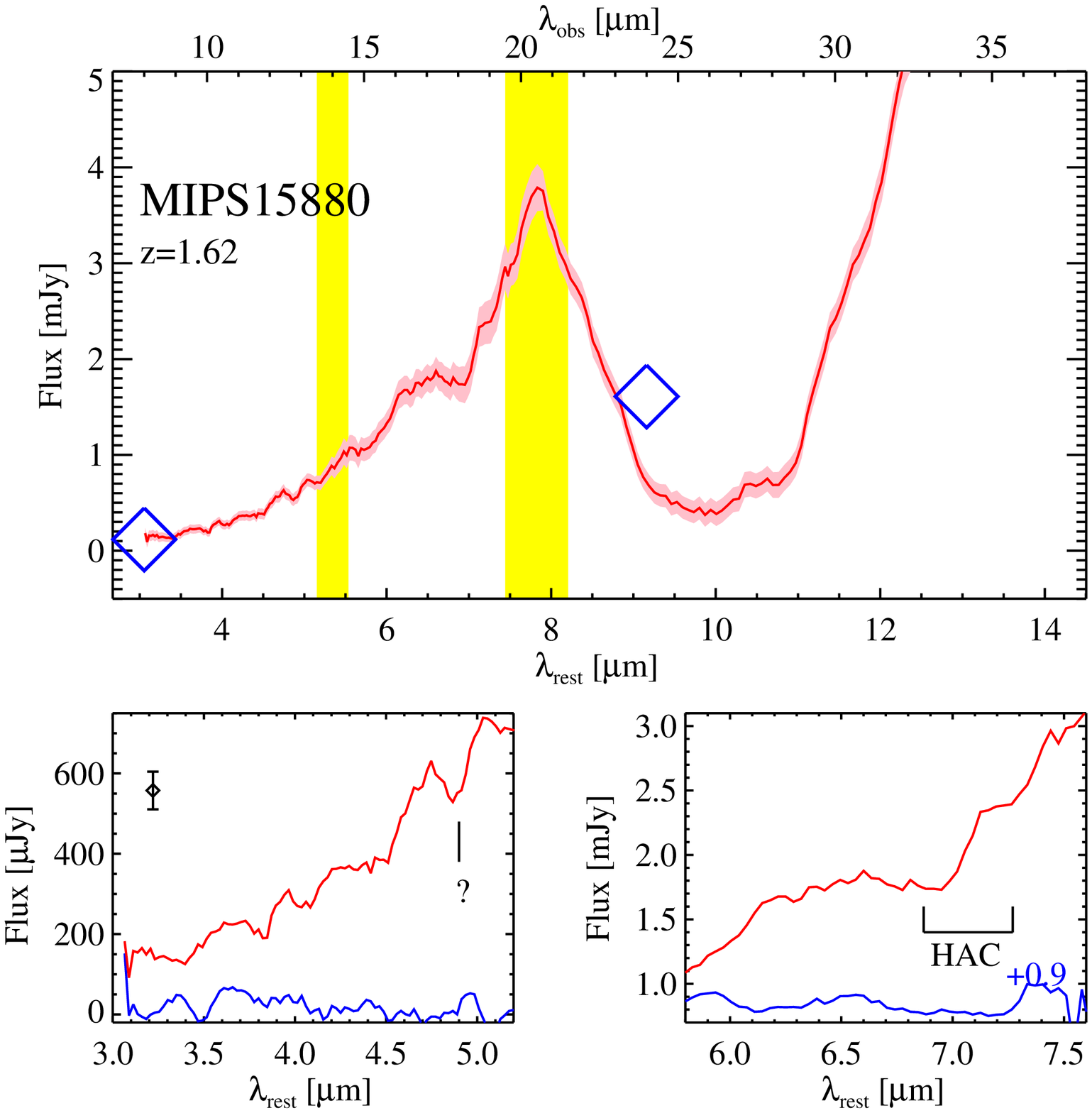}
\includegraphics[height=6cm]{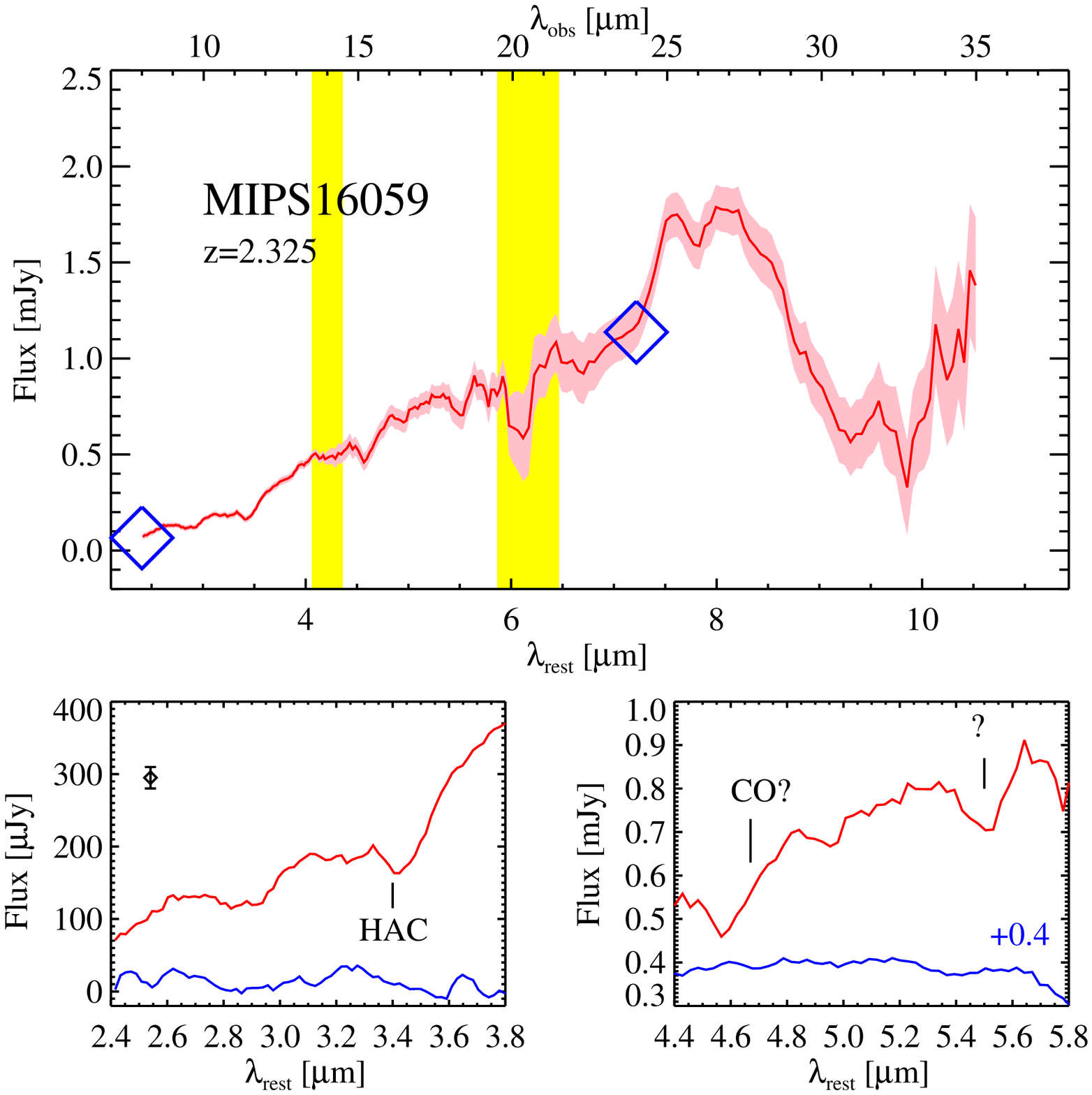}
\includegraphics[height=6cm]{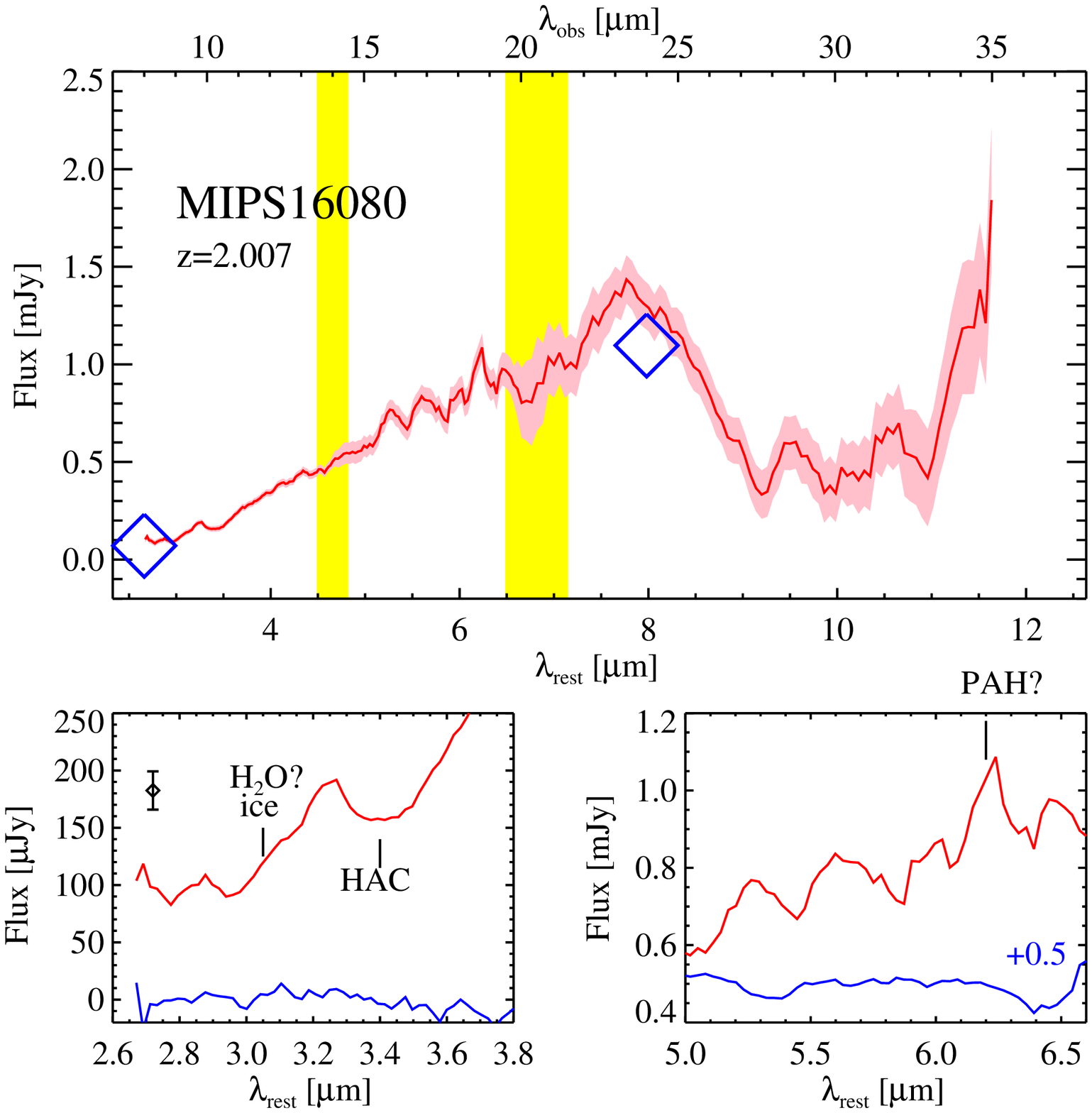}
\includegraphics[height=6cm]{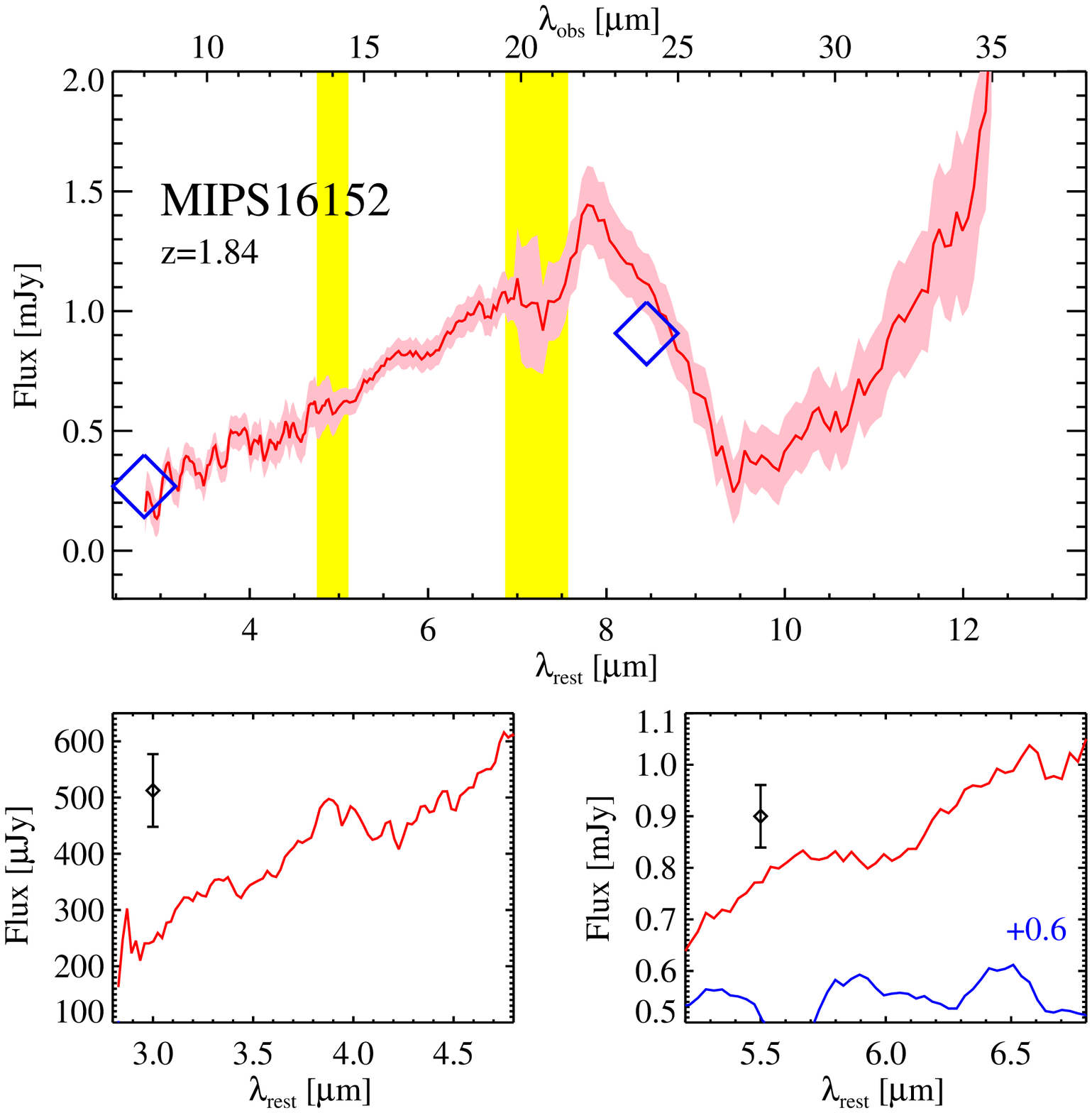}
\caption{The combined $\sim$\,2\,--\,12\um\ final spectra for our sample. The yellow bands denote the border regions where the different orders are stitched and hence any `features' that appear in those bands are likely artefacts. A smoothing of 5 pixels is applied to all spectra. The pink shaded region denotes the 1\,$\sigma$/pixel uncertainty in the spectra. The blue diamonds are the IRAC8\um\ and MIPS24\um\ broadband fluxes respectively. The two panels below are close-ups of the regions around 3\um\ and 6\um, where the errorbar denotes the mean 1\,$\sigma$ error. A sky spectrum is also shown (blue line). For easier comparison, this is often offset ($F_{\rm{sky}}$\,+$const.$), where the constant in given in blue just above the sky spectrum. }  
\label{fig_specs}
\end{figure*}

\setcounter{figure}{0}

\begin{figure*}[t]
\centering \leavevmode
\includegraphics[height=6cm]{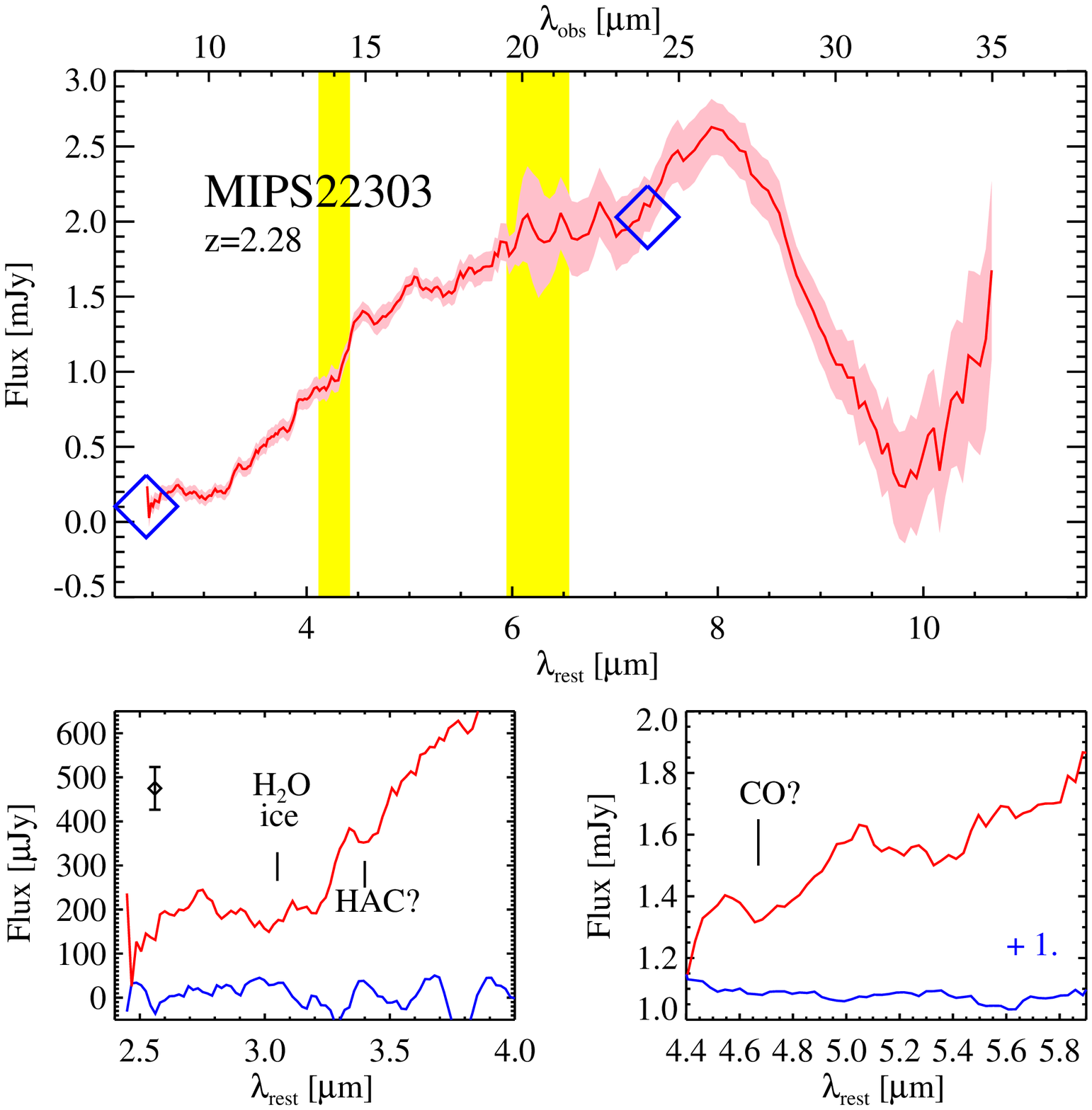}
\includegraphics[height=6cm]{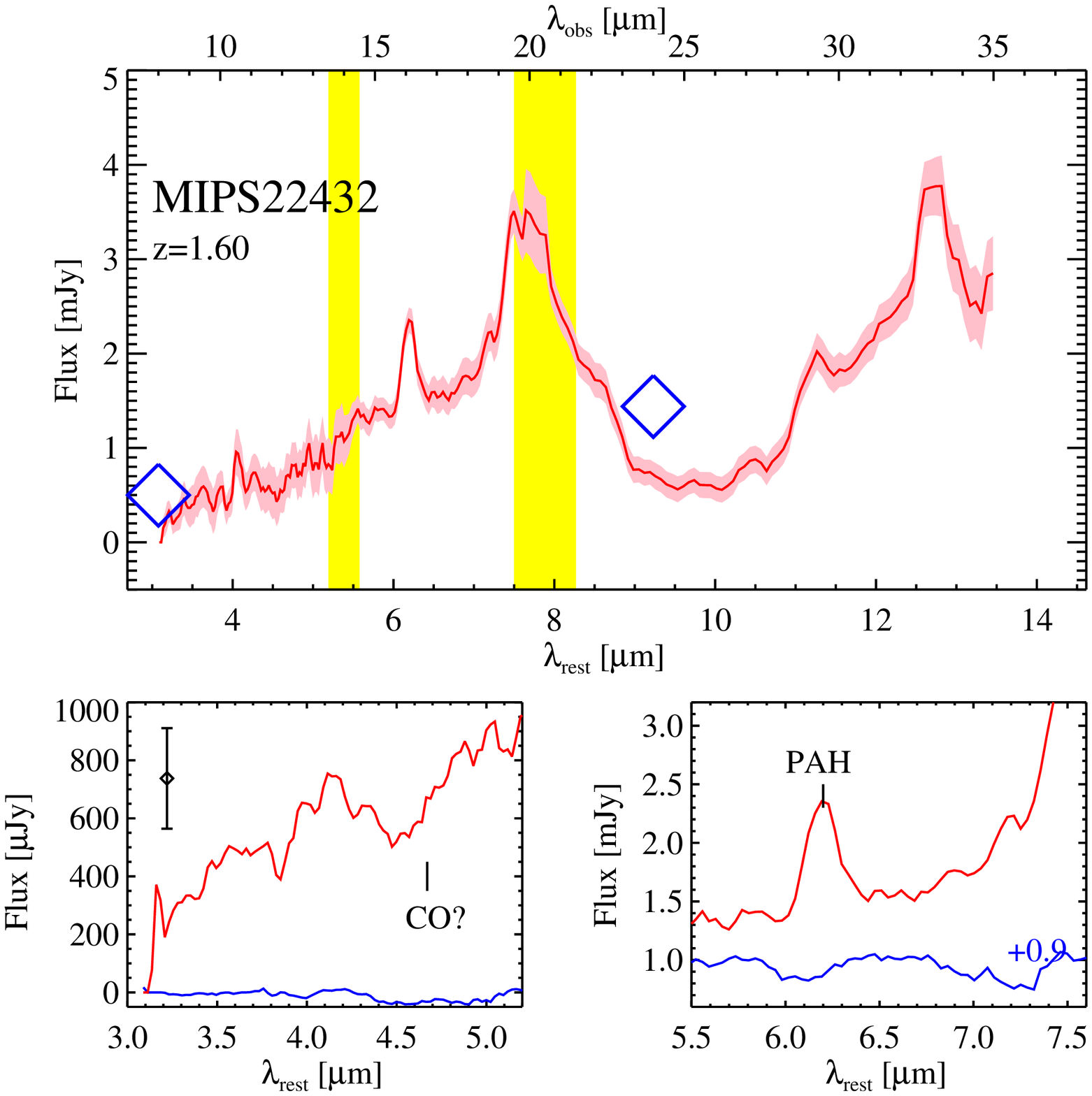}
\includegraphics[height=6cm]{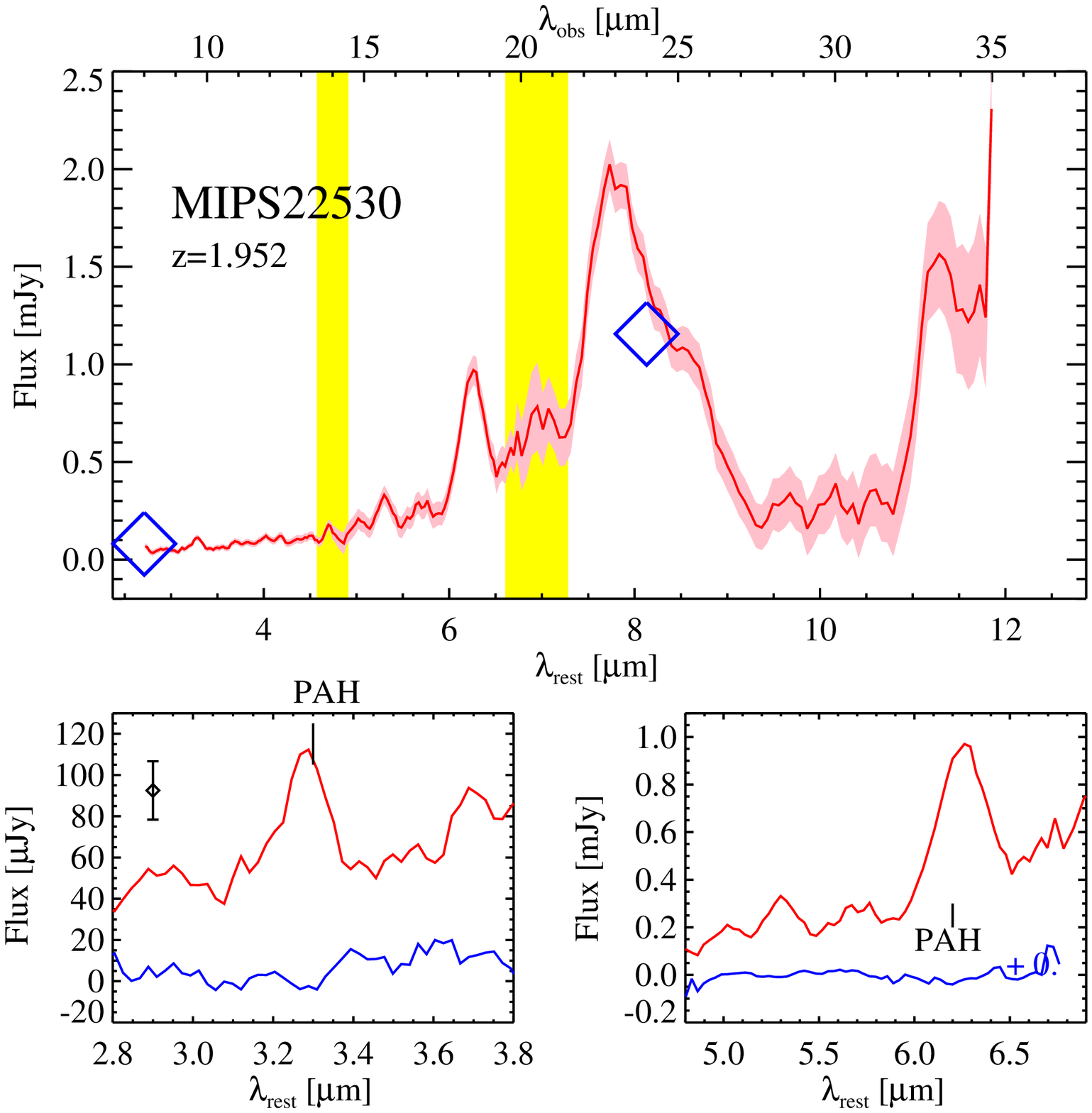}
\includegraphics[height=6cm]{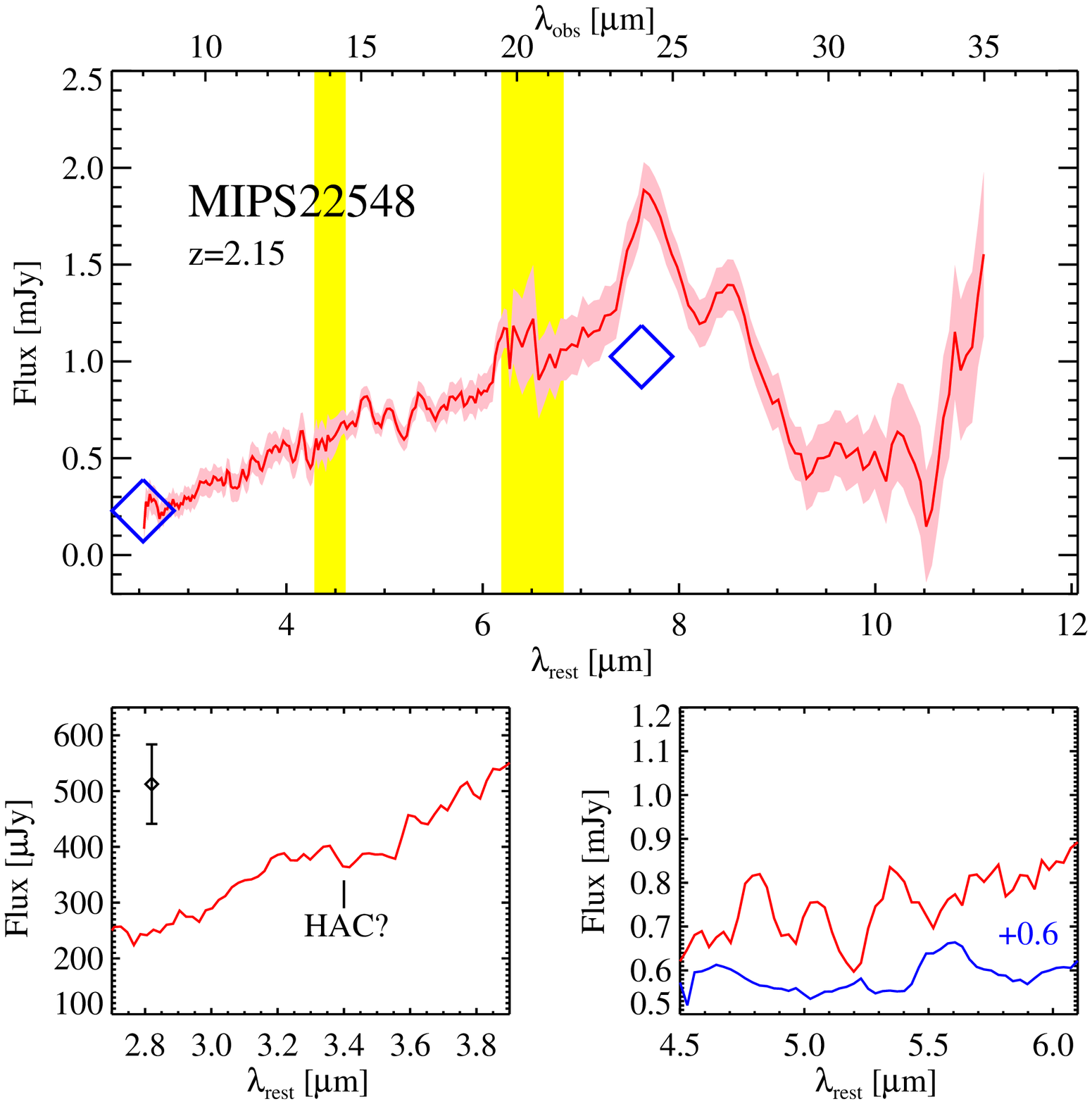}
\includegraphics[height=6cm]{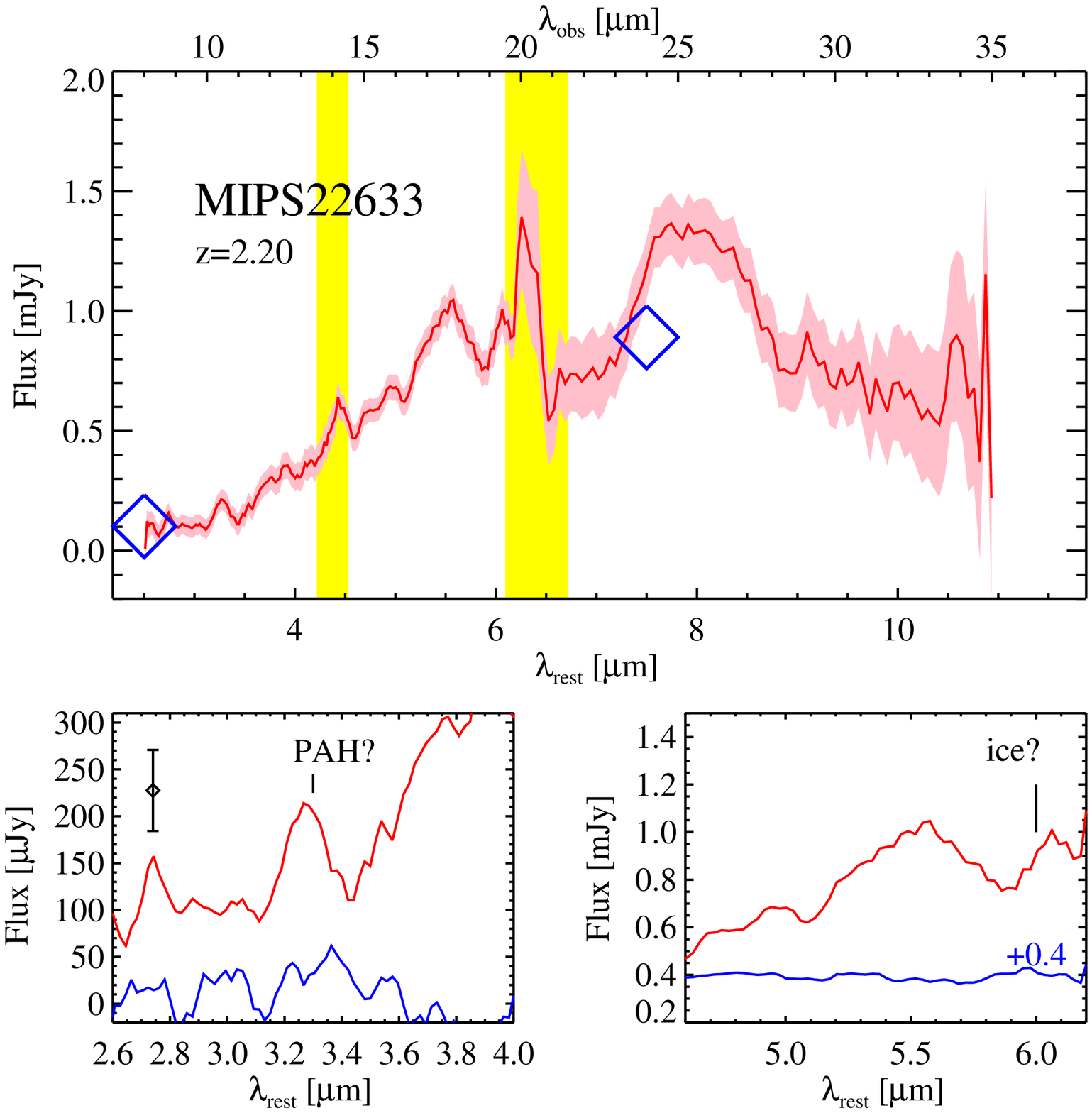}
\caption{{\it continued.}}  
\end{figure*}

\clearpage

\begin{figure}[h!]
\centering
\includegraphics[height=14cm]{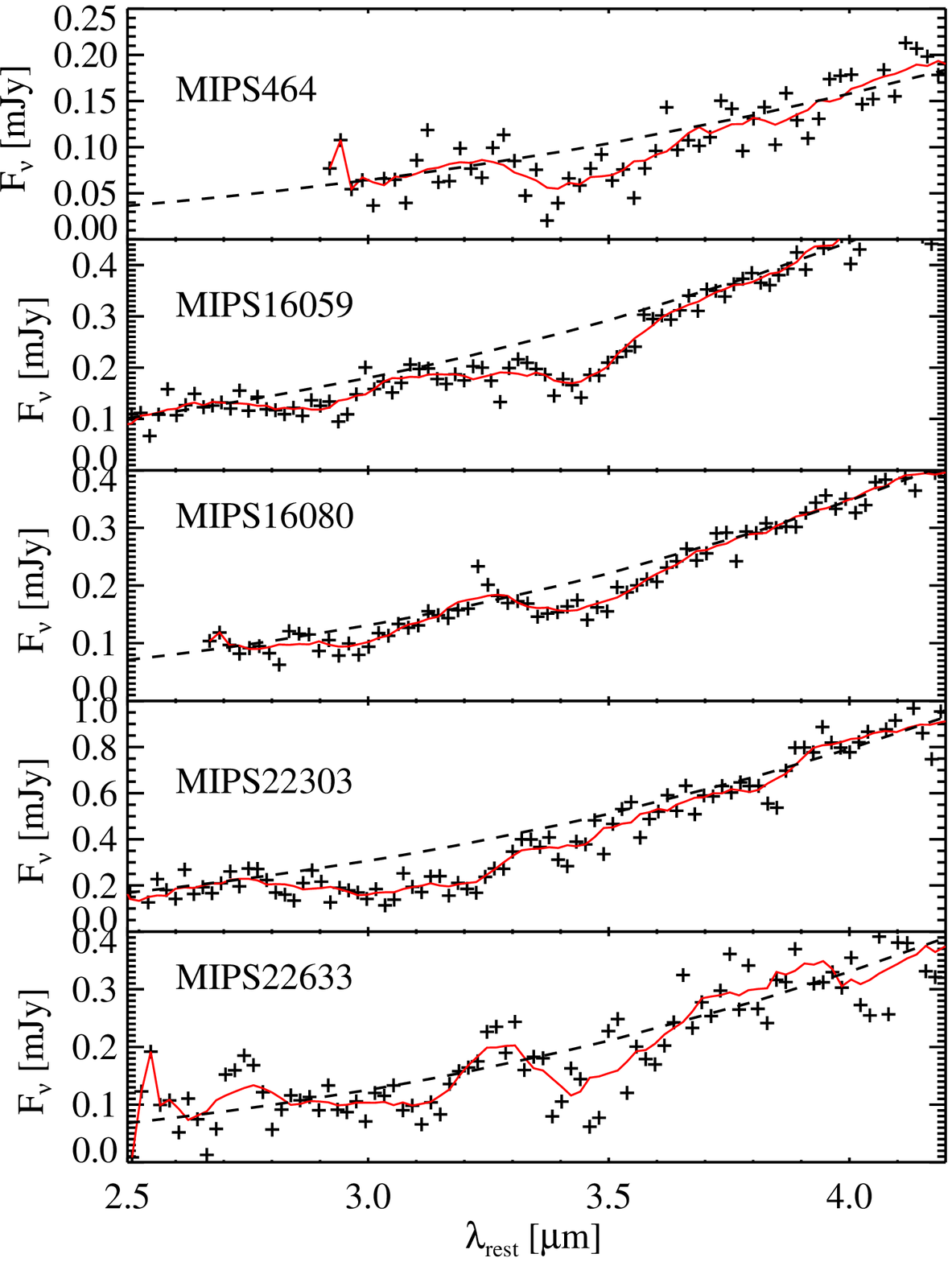}
\caption{Examples of continuum fitting in the 3\,--\,4\um\ regime. Here we show all sources showing evidence of features in this regime.  The crosses represent the actual spectra, while the solid red line is the spectra smoothed by a factor of 5. The dashed line is the best-fit continuum. \label{buried_conts}}  
\end{figure}

\begin{figure}[h!]
\plotone{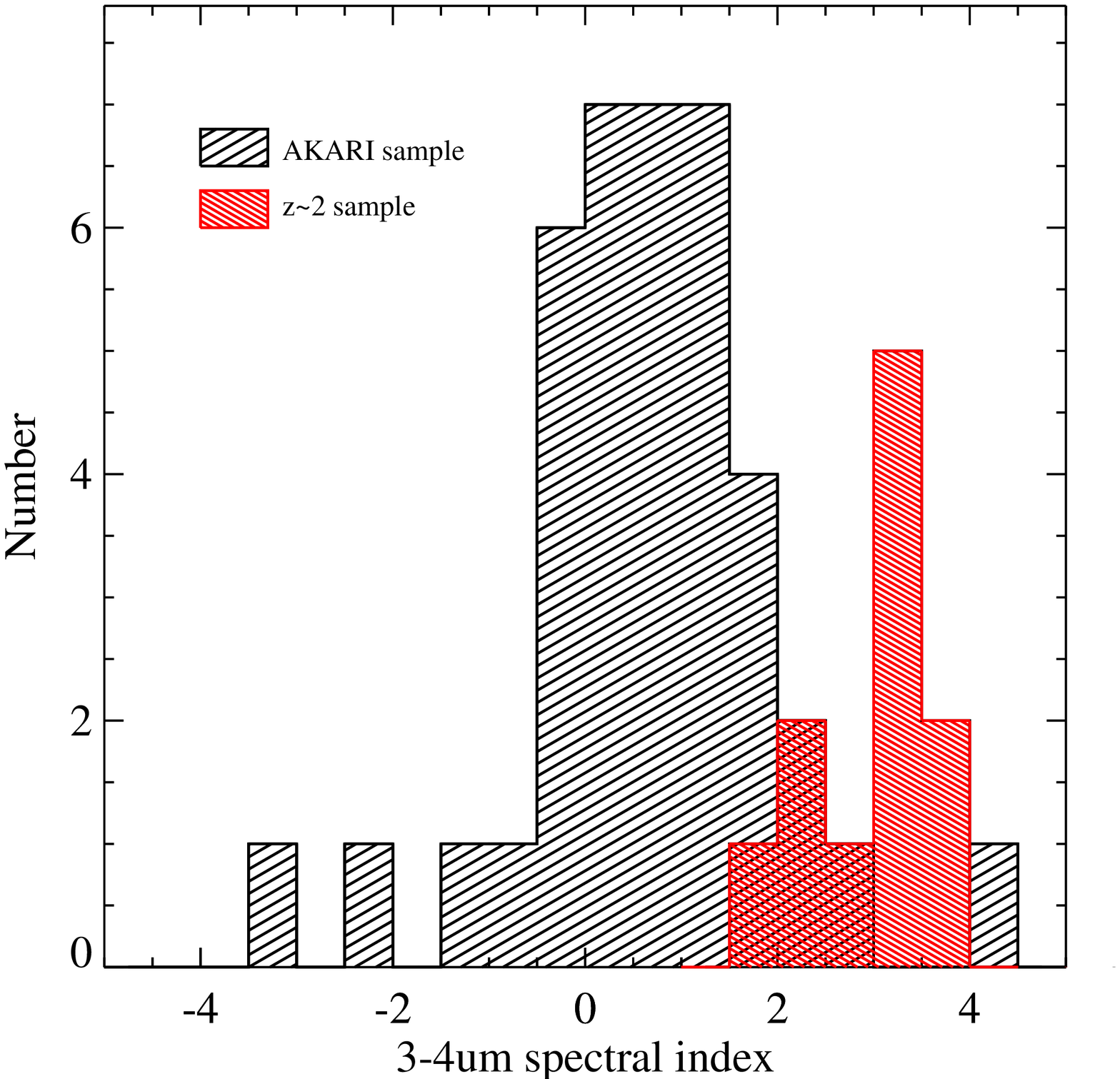}
\caption{The 3\,--\,4\um\ spectral index (defined as $f_{\nu}$\,$\propto$\,$\nu^{-\alpha}$) for our sample and the AKARI ULIRG sample of \citet{imanishi08}. Many of our sources show steeper slopes than observed locally.  \label{fig_cont}}  
\end{figure}

\subsection{Continuum slopes \label{sec_cont}}

We can see in Figure\,\ref{fig_specs} that there is considerable variety in the continuum shapes for our sources. This likely is indicative of differences in obscuration geometry. To address this fully requires comparison with radiative transfer power source+obscuration models which is beyond the scope of this paper. Here we merely wish to quantify the basic characteristics of these spectra such as spectral slope and monochromatic luminosity and compare with those of local ULIRGs. 

We adopt the standard measure of spectral slope which is a power law spectral index (i.e. $F_{\nu}$\,$\propto$\,$\nu^{-\alpha}$ or $\lambda^{\alpha}$).  We define continuum regions free of major features (2.6\,$<$\,$\lambda$\,$<$\,2.8\um\ and 3.8\,$<$\,$\lambda$\,$<$\,4.2\um). We fit the above model to the spectra that fall within the continuum regions (see Figure\,\ref{buried_conts}). For MIPS464, MIPS15880, and MIPS22432, their lower redshift means that the spectra do not reach rest-frame 2.8\um. For these sources we adopt $\lambda$\,$<$\,3.2\um\ as the continuum region. This assumes there is no water ice absorption, which is not observed for MIPS464 and MIPS16152 and is not possible to constrain for MIPS15880 or MIPS22432. We find fairly red spectra in this regime with slopes in the range $\alpha$\,$\sim$\,2\,--\,4. 

To compare our values with those of local ULIRGs without the bias of different continuum definitions, we take the ULIRG AKARI spectra from \citet{imanishi08} and apply the same continuum fitting to them.  Figure\,\ref{fig_cont} shows the distribution of the 3\,--\,4\um\ slope for our sample compared with local ULIRGs. We find that the bulk of our sample has steeper slopes than the local ULIRGs. \citet{risaliti06} argue that a red 3\,--\,4\um\ slope is an indicator of an AGN dominant in the mid-IR (although not necessarily bolometrically dominant). As defined, $\alpha$ is related to the slope $\Gamma$ as defined by \citet{risaliti06} via $\Gamma$\,=\,$\alpha$\,-2. Using this we can translate the \citet{risaliti06} criterion to: $\alpha$\,$>$\,2 sources are predominantly  AGN-dominated. This is satisfied by most of our sources, which is not surprising as most are already believed to be AGN-dominated \citep{sajina08}. MIPS22530 is our strongest-PAH, starburst-dominated source and it has the flattest spectrum with $\alpha$\,=\,1.9.

\begin{figure}[h!]
\centering
\includegraphics[height=14cm]{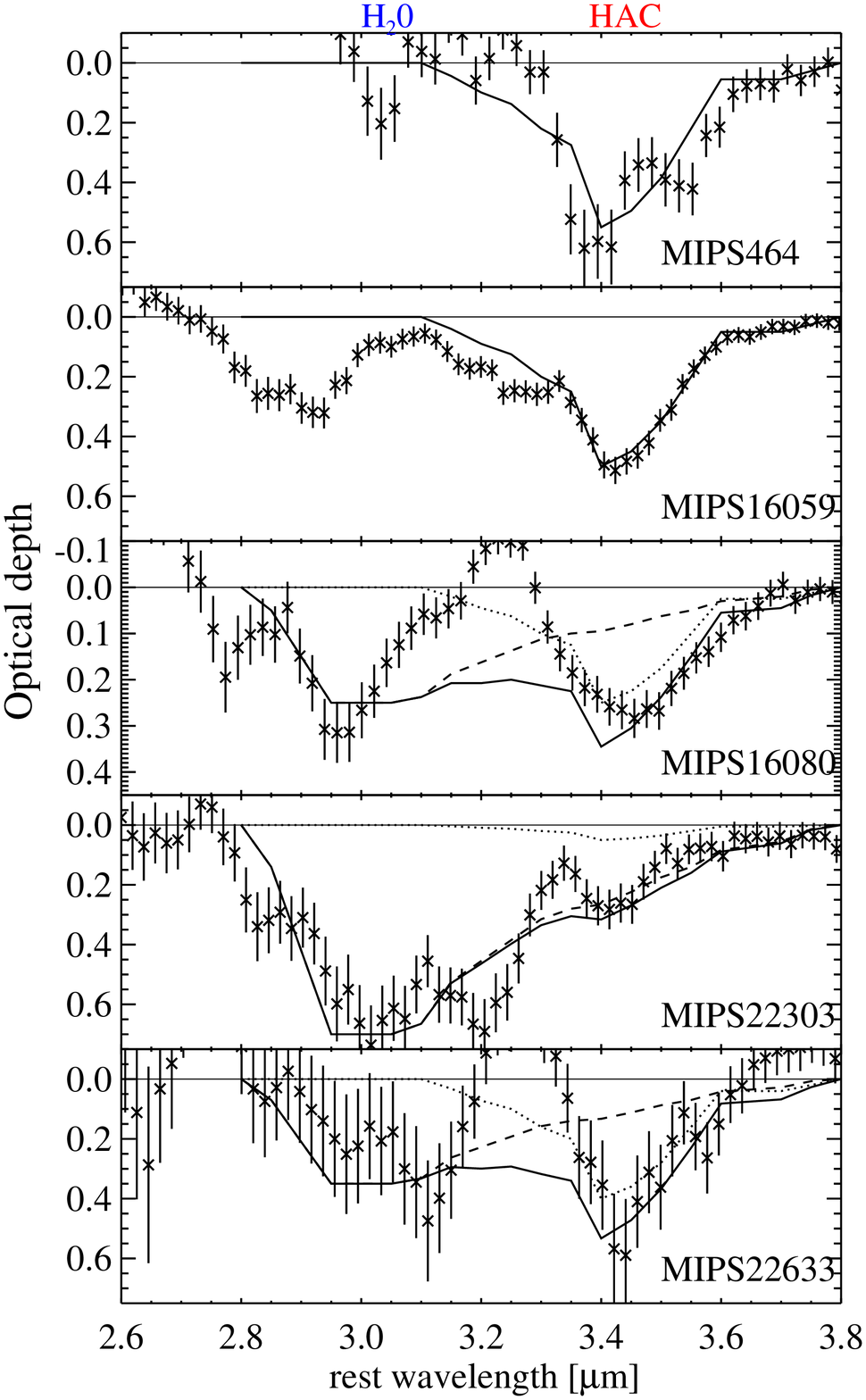}
\caption{The 3.05\um\ water ice and 3.4\um\ HAC opacity for the sources where these are observed.  We have applied a binning of 5 pixels to all spectra to increase the clarity and effective signal-to-noise ($rms/\sqrt{N_{\rm{bin}}}$).  The dashed and dotted lines are the average profiles of the 3.05\um\ water ice and 3.4\um\ HAC features for Galactic center stars from \citet{chiar02} and the solid curve is the sum of the two. Note that these are composite features and hence a wide range of specific shapes is expected \citep[e.g.][]{chiar02,duley05}. MIPS16080 and MIPS22633 show evidence of both ice and HAC features but the combined profile is inconsistent with expectations if these are the only contributors. The observed spectra could be explained by the presence of the PAH 3.3\um\ emission feature as well \citep[e.g.][]{imanishi_ulirgs}. \label{hac_det}}  
\end{figure}

We follow the same procedure for the rest-frame $\sim$\,5\,--\,7\um\ spectra, where we define as continuum regions 5.0\,$<$\,$\lambda$\,$<$\,5.5\um\ and 6.7\,$<$\,$\lambda$\,$<$\,7.0\um. The slopes there tend to be considerably flatter (except where strong-PAH contaminate this slope) as can already be seen visually in Figure\,\ref{fig_specs}. Lastly, we compute the monochromatic 3\um\ and 5.5\um\ luminosities based on these fits. Table\,\ref{table_slopes} gives the continuum parameters (luminosites and slopes) of our sources. 

\subsection{Optical depth measurements: feature reliability \label{sec_optdepth}}

In the previous section, we discussed the determination of the continuum levels for our sources. Figure\,\ref{buried_conts} shows the best-fit continua for the sources with water ice and/or 3.4\um\ HAC absorption (MIPS464, MIPS16059, MIPS16080, MIPS22303, and MIPS22633). This figure also shows that the absorption features are not driven by smoothing over a single outlier/spike. None of the sky spectra of these sources show broad features of such strength which also excludes them being background subtraction artefacts. After the continuum fitting, the optical depth of the various features is then determined based on: $\tau$\,=\,ln$(F_{\rm{cont}}/F_{\rm{data}})$.  We should stress that here we use the convention of calling this observed feature strength an optical depth, even though the true optical depth is dependent on the exact radiative transfer solution. In particular, in the case of the realistic clumpy dust distribution the two do not translate in any straightforward way \citep{nenkova08}. As discussed previously, the reliability of weakly detected features is complicated by the frequently uncertain redshifts (except for MIPS16059, MIPS16080, and MIPS22530 which sources have Keck rest-frame optical spectra). For MIPS464, MIPS22303, and MIPS22633, the observed features are of a redshift consistent with the silicate feature and sufficiently strong in themselves to make the feature identification convincing.  Table\,\ref{table_taus} gives a summary of the detected features as well as the updated redshifts. The uncertainties are estimated based on $F_{\rm{data}}$\,$\pm$\,1$\sigma$. We ignore the uncertainty in the continuum level.

\subsection{Water ice and HAC absorption \label{sec_ice}}

\begin{figure}[h!]
\centering
\includegraphics[height=10cm]{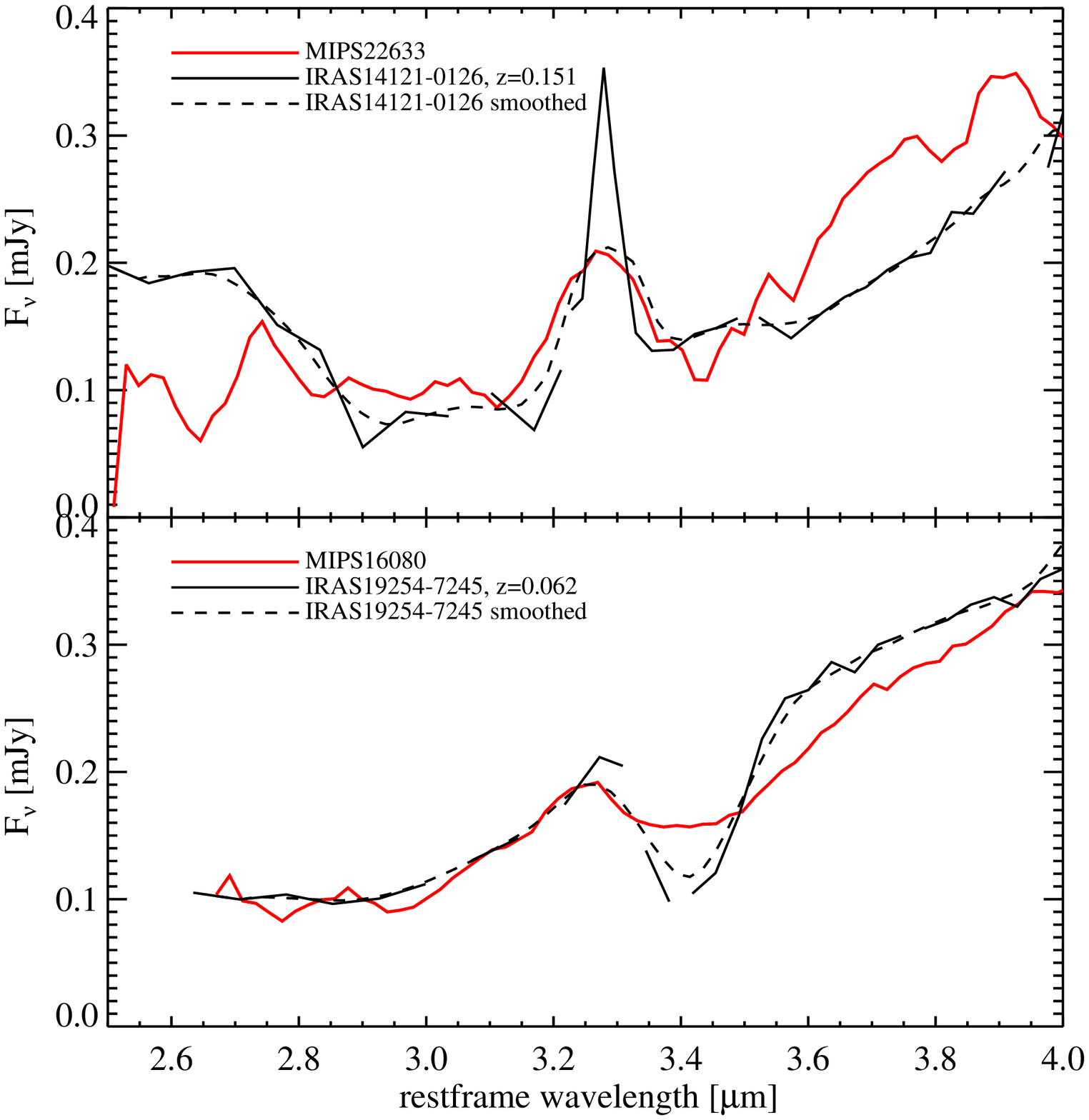}
\caption{The spectrum of MIPS22633 compared with the AKARI spectrum of IRAS14121-0126 ($z$\,=\,0.151). The later has been degraded to the spectral resolution of the MIPS22633 spectrum. The combination 3.3\um\ PAH+3.05\um\ water ice absorption for MIPS22633 is quite feasible. \label{akari_comparison}}  
\end{figure}

In \S\,\ref{sec_optdepth} we claim detections of water ice and/or HAC absorption for  5/11 of the sources in our sample (see Table~\ref{table_taus}).  Another test of the reliability of these features is to compare their observed profiles with the average profiles of these features as determined for Galactic sources.  In Figure\,\ref{hac_det} we show the same five sources as in Figure\,\ref{buried_conts} but with the continuum subtracted, and the average profiles of the 3.05\um\ water ice and 3.4\um\ HAC features from \citet{chiar02} overlaid. The continuum is the same as determined in the previous section, although it should be kept in mind that there is some uncertainty in the continuum-fit parameters, as the fit is done on fairly small sections of essentially feature-free spectra. We find that there is broad agreement although our spectra show noise about these profiles.  The $\sim$\,3.0\um\ absorption feature in MIPS16059 is at somewhat lower wavelengths than expected for water ice and therefore we treat this as an upper limit only and not as a detection. Overall, the most significant deviation is observed for MIPS16080, MIPS22303 and MIPS22633 all of which show an excess at 3.3\um\ in addition to the likely water ice and/or 3.4\um\ HAC absorption. To test the possibility of this being due to the 3.3\um\ PAH, in Figure~\ref{akari_comparison}, we compare the MIPS22633 and MIPS16080 spectra with reasonably closely matching local ULIRGs where we degrade the resolution of the AKARI spectra to match that observed. Both local comparison sources show 3.3\um\ PAH \citep{imanishi08}, although in IRAS 19254-7245 it is much weaker than in IRAS14121-0126. We find that indeed the 3.3\um\ excess can very well be due to PAH emission especially for MIPS22633. For MIPS16080 the comparison is less convincing and therefore we treat its 3.3\um\ emission as an upper limit only. 

\begin{figure}[h!]
\plotone{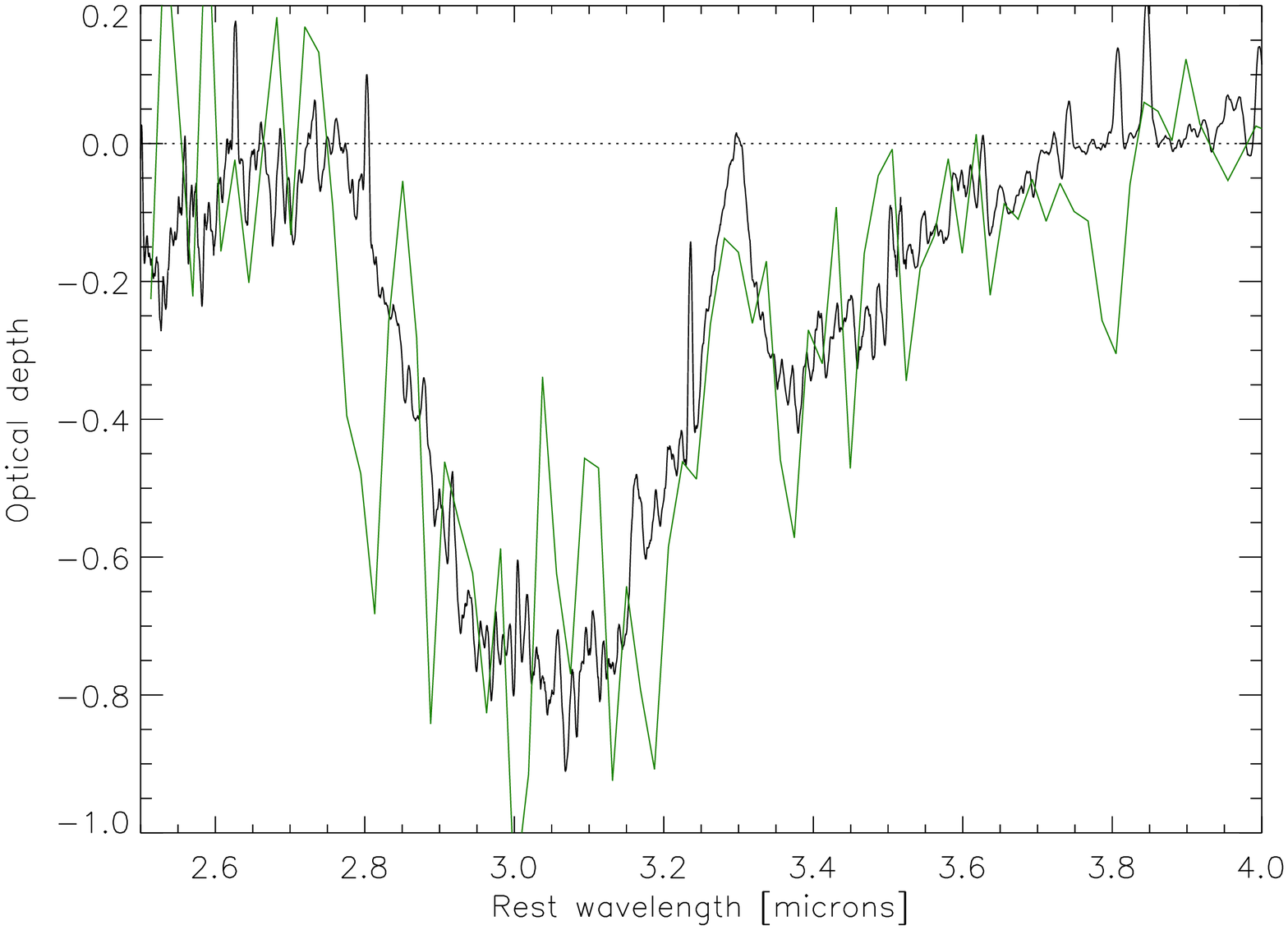}
\caption{ The unsmoothed spectrum of MIPS22303 (green curve) compared with the ISO-SWS-01 spectrum of the north eastern part of the molecular ring around the Galactic Center (black curve). For MIPS22303, the combination water ice and 3.3\um\ PAH without HAC absorption seems quite likely. \label{fig_mips22303}}  
\end{figure}

For MIPS22303 we find an excellent local analog in the {\sl ISO}-SWS spectrum of the Galactic Center source GC Ring NE (see Figure\,\ref{fig_mips22303}). This comparison suggests, that MIPS22303 is well described by a combination water ice+3.3\um\ PAH without any clear need for HAC absorption (as shown in Figure\,\ref{hac_det}). For this source therefore, we only have an upper limit on the HAC absorption. 

\begin{figure}[h!]
\plotone{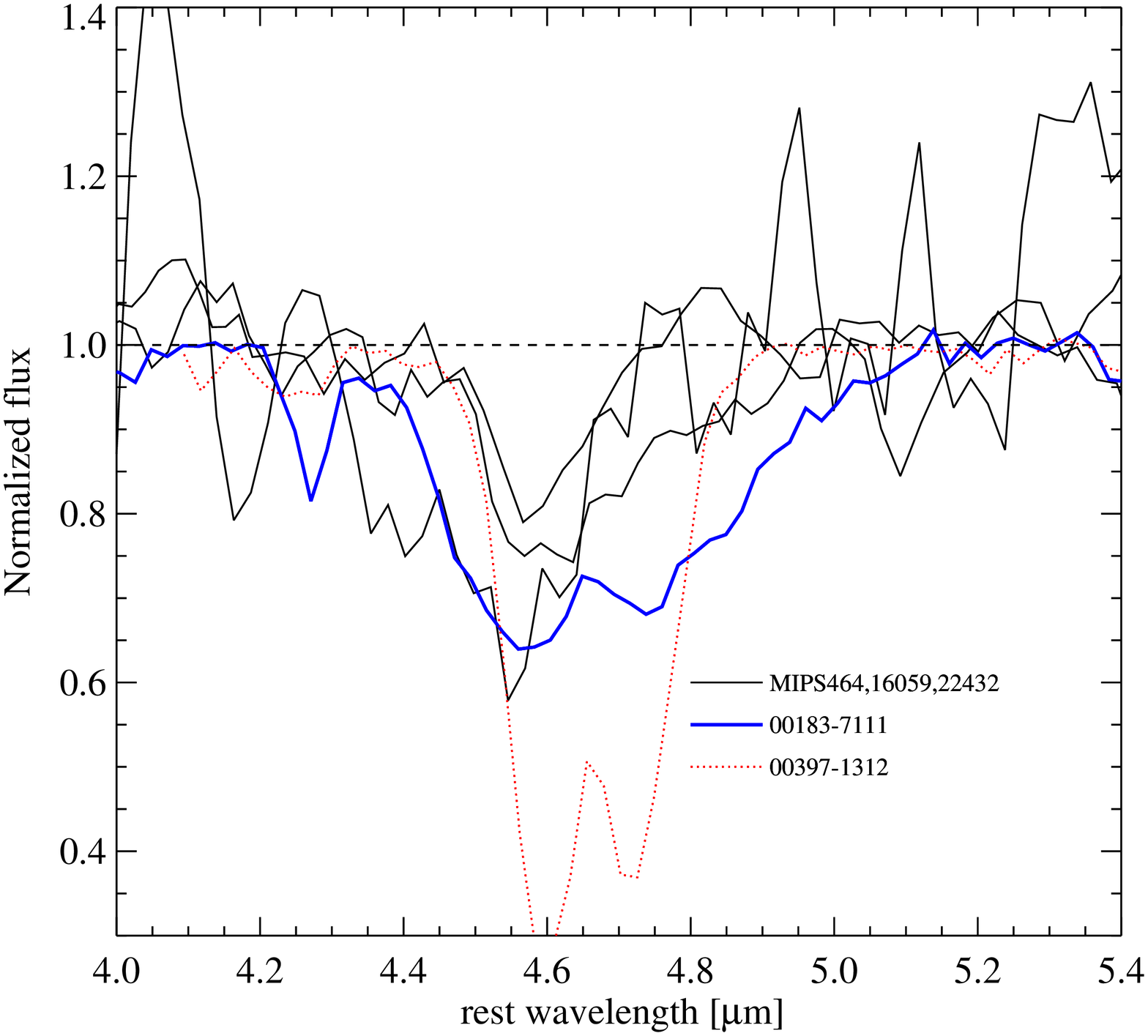}
\caption{The three sources with signs of CO absorption. The $y$-axis shows normalized flux (i.e. $\tau$\,=$\ln$(normalized flux)). For comparison we overlay the same for 00183-7111 and 00397-1312.  The noisiest spectrum of our sources is that of MIPS22432 which is responsible for the large spikes observed.  \label{fig_co}}  
\end{figure}

\subsection{4.67\um\ CO absorption \label{sec_co}}

One of the goals of this program was to try to detect the warm CO gas absorption feature at 4.67\um. Unfortunately, in several cases, rest-frame 4.67\um\ falls right in the region of where we stitch together the SL1 and LL2 orders (yellow band in Figure~\ref{fig_specs}) making any features observed there unreliable. Sources with $z$\,$>$\,2.1 (i.e. where this region is entirely contained in the LL2 spectra) or with $z$\,$<$\,1.8 (i.e. where this region is entirely contained in the SL1 spectra) are the only feasible ones to look at. There are a total of 7 sources that satisfy these criteria. Of these, three sources show signs of absorption in this regime: MIPS464, MIPS16059, and MIPS22432 (see Figure\,\ref{fig_specs}). 
These features are fairly convincing in themselves (compared with the background), but do they agree with the  expectations for the CO feature? In Figure\,\ref{fig_co} we compare their smoothed normalized flux spectra (i.e. data/cont), with the same for the hyperluminous galaxy F00183-7111 and the IRAS00397-1312 \citep{spoon_co}. While the nominal position of the line is at 4.67\um, the peak absorption in our sources (assuming it is CO) is typically blueward of that. As the CO gas absorption feature is composed of two broad branches (R and P), if one of these is stronger than the other it results in the observed shift. A stronger R branch (shorter-$\lambda$) absorption is also observed in IRAS08572+3915 \citep{imanishi08} and IRAS00397-1312 \citep{spoon_co,imanishi09}. Note that for all these galaxies, the redshift is well established on the basis of other features, without any additional shifts required to match the CO feature. For MIPS464 the redshift is based on  the silicate absorption plus 3.4\um\ HAC feature. For MIPS16059, the redshift is based on Keck spectra, but is also consistent with the silicate and 3.4\um\ features.  For MIPS22432, the only strong-PAH source of the three, the redshift is based on the 6.2\um\ PAH feature.  

The observed strength of the CO feature for these three sources (the strongest in our sample) is $\tau_{4.6}$\,$\sim$\,0.2 (as discussed in \S\,4.3 this is just the observed feature strength and does not translate to intrinsic optical depth). This is somewhat weak compared with the CO feature strengths observed by \citet{sani08}, which range from $\sim$\,0.1\,--\,2.2.  The relative weakness of our features can also be seen in Figure\,\ref{fig_co}. 

An possible explanation for both the observed asymmetry and the overall weakness of the CO feature might be the effects of a starburst-powered continuum originating from material outside the CO absorption layer. MIPS222432 (a strong-PAH source) has a clear starburst-component and as we show in \S\,4.8, a starburst component is likely present in all sources even if not individually observed. 

\subsection{PAH features}

We look for the 3.3\um\ and 6.2\um\ PAH features which are not as confused with the continuum model as for e.g. the strongest PAH feature at 7.7\um\ \citep{sajina07}, and hence might reveal the suspected presence of star-formation even in sources dominated by nuclear activity. In \citep{sajina07} we showed that the stacked spectra of weak-PAH (based on the shallower spectra) sources show PAH emission. There are several cases of weak-PAH sources that have cold dust detections \citep{sajina08}.  In addition two our sources, MIPS22432 and MIPS22530 were identified as strong PAH sources based on our previously defined criteria \citep{sajina07}. MIPS22530 was also shown to be a starburst-dominated source based on its rest-frame optical spectrum and cold dust detection \citep{sajina08}.  

MIPS22432 already had the necessary signal-to-noise ratio and hence no new data for it is presented here. We merely re-reduced the spectra and again see a clear 6.2\um\ PAH feature. Unfortunately, the 3.3\um\ feature falls right on the edge of its spectrum and hence we cannot test for its detectability.  

MIPS22530, on the other hand, shows an unambiguous 3.3\um\ PAH feature (see Figure\,\ref{fig_specs}). In addition, three more sources MIPS16080, MIPS22303 and MIPS22633 show some evidence for the 3.3\um\ PAH feature but somewhat confused with water ice and HAC absorption (see Figure,\ref{hac_det}).  For MIPS22303 and MIPS22633, a potential 6.2\um\ feature falls in the unreliable wavelength range where we stich the different IRS orders (see Figure\,\ref{fig_specs}). MIPS16080, whose 3.3\um\ PAH feature is too weak to be considered a detection, does show a tentative detection of the 6.2\um\ feature. The PAH emission of this source is made more reliable due to its having an accurate optical spectroscopic redshift. In addition, MIPS16080 has starburst-like optical line ratios, but AGN-line IR spectrum suggesting it is a composite source \citep{sajina08}.

\begin{figure}[h!]
\plotone{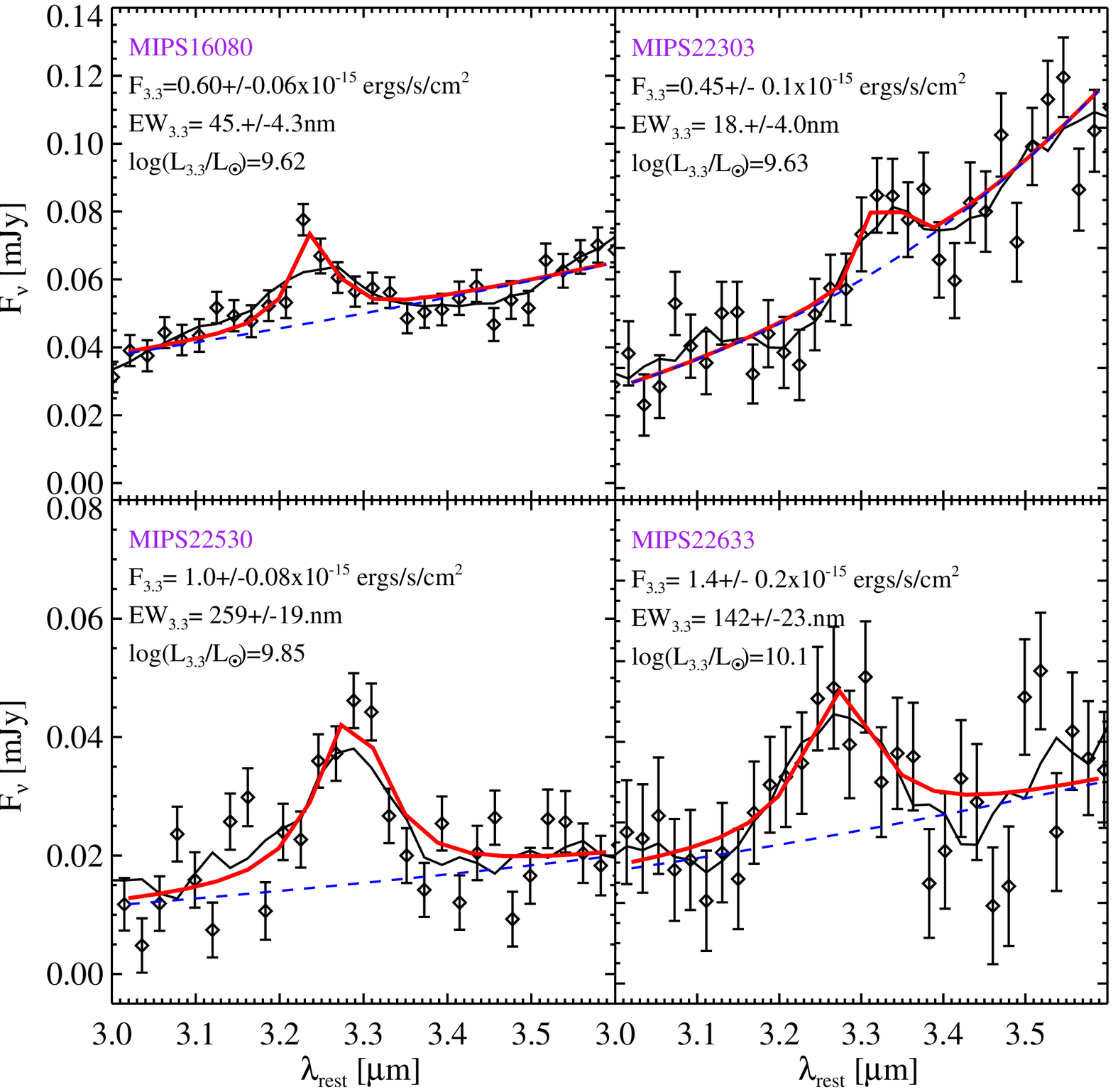}
\caption{The fits to the 3.3\um\ PAH emission for the four sources that show signs of it this feature. The data points show the spectra themselves, the solid black lines show the smoothed spectra. The best-fit continuum is given by the blue dashed line, whereas the total best-fit model is given by the thick red line. \label{pahfit_3um}}  
\end{figure}
 
We fit a simple local continuum+Lorentz profile model to each of the detected or hinted at 3.3\um\ PAH features. Figure\,\ref{pahfit_3um} shows the results.   Wherever it was possible, we did the same for the 6.2\um\ feature.  For MIPS16080 we find $f_{6.2}$\,$\simeq$\,2.5\,$\times$\,$10^{-15}$erg/s/cm$^2$ (EW\,$\simeq$\,120\,nm). For MIPS22530 we find $f_{3.3}$\,$\simeq$\,0.97\,$\times$\,$10^{-15}$erg/s/cm$^2$ (EW\,$\simeq$\,550nm)  and  $f_{6.2}$\,$\simeq$\,3.89\,$\times$\,$10^{-15}$erg/s/cm$^2$ (EW\,$\simeq$\,370\,nm). These imply a 6.2/3.3 ratio of $\sim$\,3\,--\,4 for these sources. This is roughly consistent with what is observed for local ULIRGs \citep[e.g.][]{imanishi_irs,imanishi08} although the observed ratio has a large scatter.  Note however, that for MIPS16080, both the 3.3 and 6.2\um\ PAH features are somewhat unreliable (see e.g. Figure\,\ref{akari_comparison}) and should really be considered upper limits. 

\begin{figure}[h!]
\centering
\includegraphics[height=14cm]{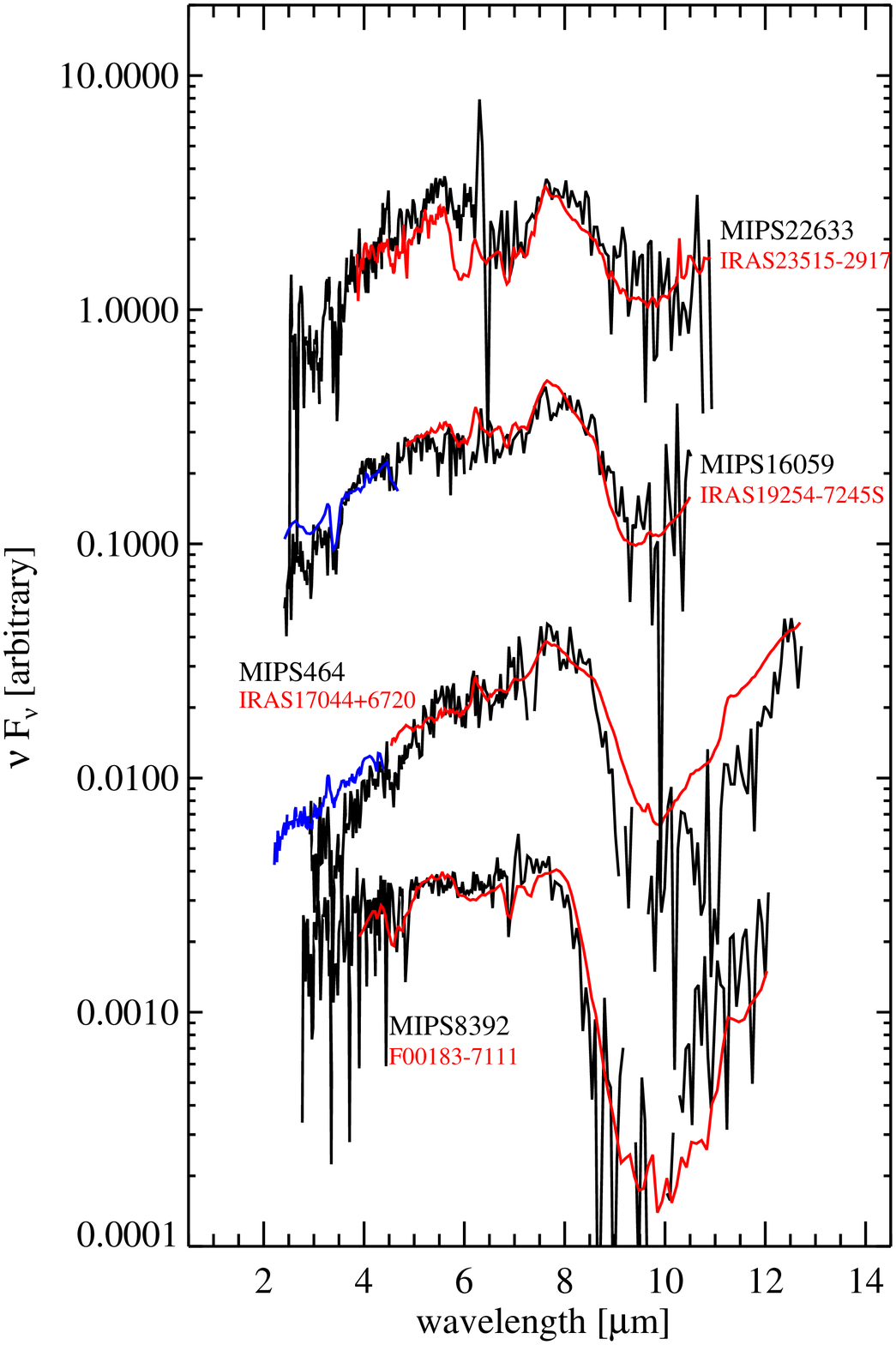}
\caption{From top to bottom, in order of increasing silicate feature depth: MIPS22633, MIPS16059, MIPS464, and MIPS8392. The red and blue curves show respectively the Spitzer IRS and AKARI (when available) spectra for local analogues. From top to bottom these are: IRAS23515-2917, IRAS19254-7245, IRAS17044+6720, and F00183-7111.   \label{fig_comp}}  
\end{figure}

\subsection{Comparison with local analogs}

Figure\,\ref{fig_comp} shows our sources overlaid with local analogs. From top to bottom we show sources in order of increasing silicate feature depth. We picked IRAS19254-7245, and IRAS17044+6720 due to their being among the few sources in the AKRI sample of \citet{imanishi08} which had comparably red 3\,--\,4\um\ spectra.  Here they are compared with MIPS16059 and MIPS464 respectively, to show that selecting on the basis of red 3\,--\,4\um\ spectra alone, yields good agreement across the full observed range ($\sim$\,2\,--\,11\um). MIPS22633 looks quite extreme and unusual, showing a very strong $\sim$\,6\um\ absorption complex, while being the weakest silicate absorption source in this sample. However, we do find a local analog, IRAS23515-2917, which has an even more extreme 6\um\ absorption coupled with a relatively shallow 9.7\um\ silicate absorption. Lastly our strongest silicate absorption source, MIPS8392 seems to have a mid-IR SED remarkably like that of the hyperluminous galaxy F00183-7111 \citep{spoon_f00183}. This galaxy is one of the very few local examples that have comparable luminosities to those of our sample, and therefore finding such good agreement is not surprising. 

\begin{figure}[h!]
\plotone{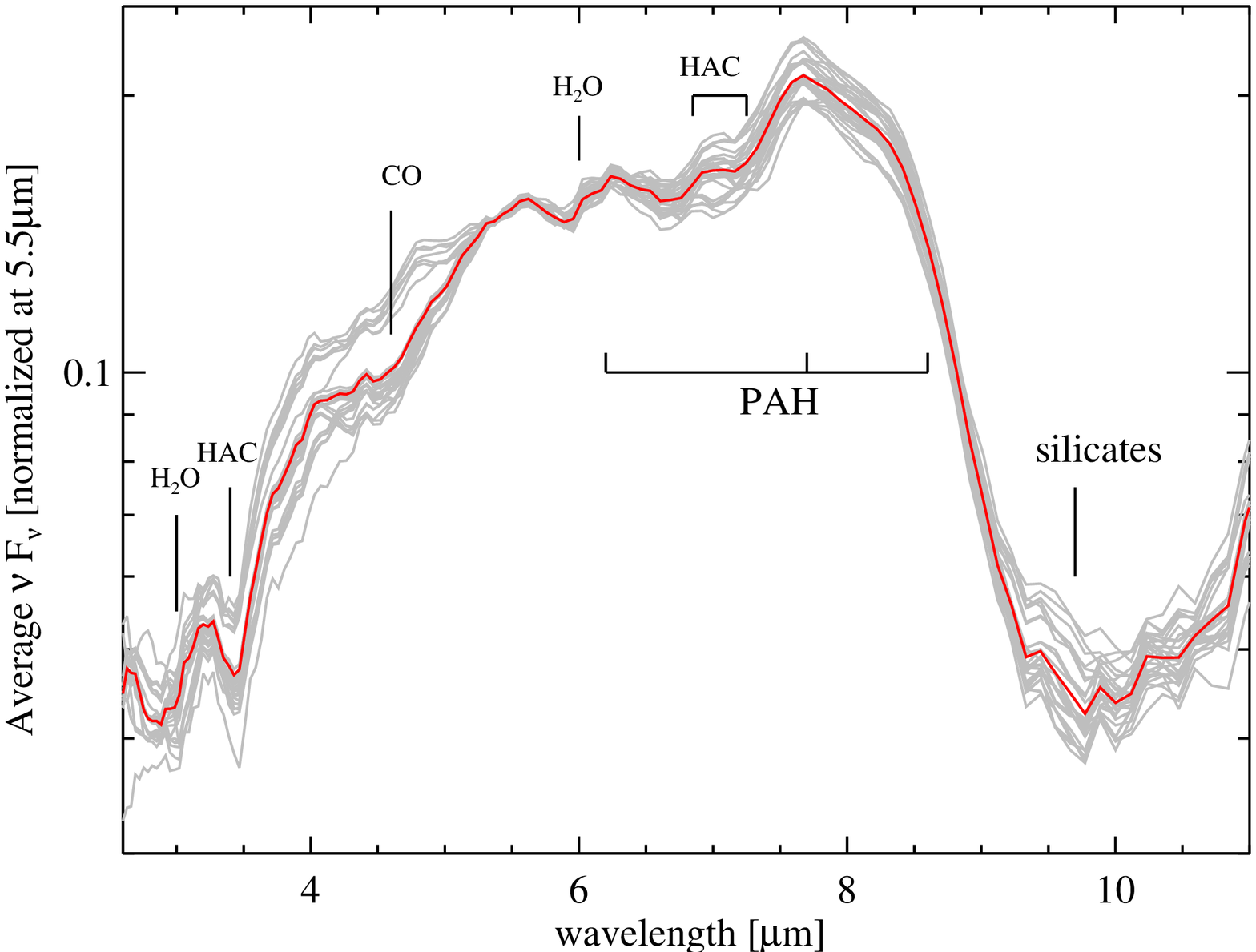}
\caption{The red curve shows the average spectrum of all sources excluding the strong-PAH sources MIPS22432, and MIPS22530 and MIPS8392 which is of too poor signal-to-noise below $\sim$\,5\um. To assess the effect of individual sources, the grey curve show average spectra containing 6/8 of the above sources. In each of the curve a different pair of sources is excluded in order to show that some aspects of the stack are there independent of which sources are included or not (see text for details). The stacks are performed on un-binned data with all individual spectra normalized at 5.5\um\ before co-adding. This point was chosen because it is in the center of our new spectra, it is pure continuum in all sources unaffected by edge effects and near the 6\um\ water ice absorption which we want to look for. The only clearly outlying set of stacks are the four ones that exclude MIPS464.   \label{fig_stack}}  
\end{figure}
 
\subsection{Average spectrum \label{sec_avspec}}

In the previous sections we have shown that a number of features are detected in the individual sources spectra including the 3.4\um\ HAC feature as well as the 3.05\um\ water ice absorption feature. There are hints of some other features in the sources such as marked in Figure\,\ref{fig_specs}, but they are typically of low significance. These include in particular the 6\um\ absorption complex (likely including the 6.15\um\ water ice feature), the 6.85 and 7.25\um\ HAC features and the 4.67\um\ CO feature discussed in \S\,\ref{sec_co}. To address the question of whether or not these are present in our sample, we stacked the spectra as shown in Figure\,\ref{fig_stack}. Because there are too few sources to test the robustness of the stack by taking randomly half of it, we used a different technique which involves removing each existing pair of sources, or 28 possibilities. This implies that features that appear in each of the 28 stacks need to be present in the majority of sources in order not to disappear in at least some of the combinations. 

\begin{figure}[h!]
\plotone{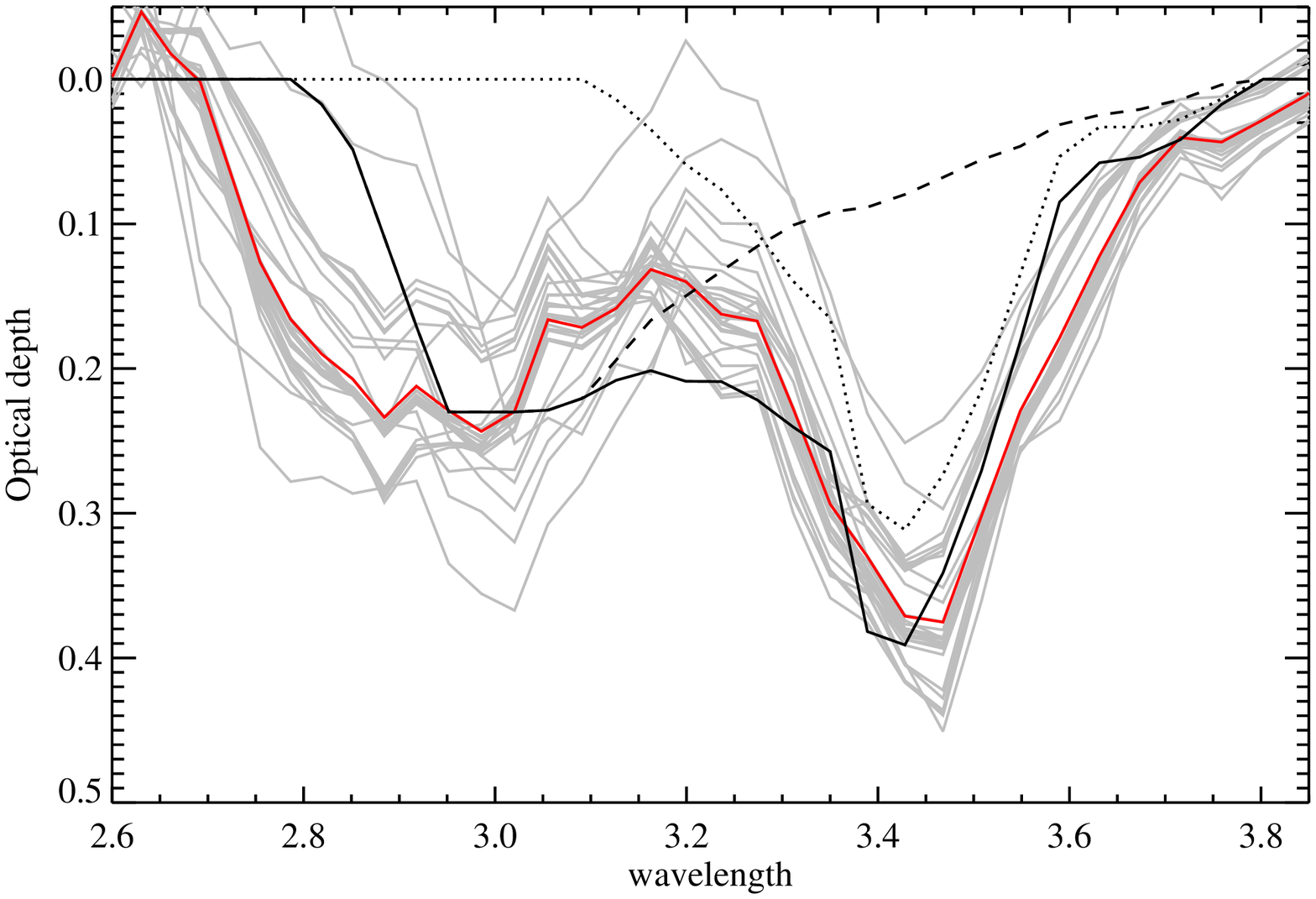}
\caption{The stacked 3\,--\,4\um\ optical depth profile overlaid with the ice and HAC profiles as in Figure\,\ref{hac_det}.  To obtain this, a continuum was fit to each stacked spectrum. As in Figure\,\ref{hac_det} the red line is the average spectrum with the grey lines being stacked spectra with pairs of sources removed. \label{fig_stackzoom}}  
\end{figure}

Figure\,\ref{fig_stackzoom} shows that in the stacked spectra, we see that the 3.4\um\ HAC feature is quite prominent, unsurprising given that we detect it in four of the individual spectra. Potentially there is also 3.05\um\ water ice absorption and as well as potentially the 3.3\um\ PAH emission given some 3.3\um\ excess. The 3.0\um\ feature is badly fit at the lower wavelength end for most, but not all permutations. MIPS16059 is probably responsible for some fraction of these (see \S\,4.4), the rest is probably due to edge effects in sources without any clear features. Figure\,\ref{fig_stack}, however, suggests that the 6\um\ absorption complex is there in most sources. The most prominent such feature is seen in MIPS22633 where the 6\um\ absorption is clear in the individual spectrum as well. One thing to keep in mind is that in a number of the sources, the $\sim$\,6\um\ features in particular are not individually detected due to their falling inside the unreliable IRS modules overlap region.  This uncertainty, as well as the likely 3.3\um\ PAH make it difficult to estimate the average 3.0\um\ water ice and 3.4\um\ HAC feature depths. Figure\,\ref{fig_stackzoom} suggests that these are $\langle\tau_{3.0}\rangle$\,$\sim$\,0.22 and $\langle\tau_{3.4}\rangle$\,$\sim$\,0.30, but it should be born in mind that these are fairly uncertain.

The stacked spectra show signs of PAH emission superimposed on the strong mid-IR continua (recall that we have excluded the two individually strong-PAH sources in the sample). The $\sim$\,8\um\ profile suggests the presence of the 7.7\um\ and 8.6\um\ features. In addition, the 3.3\um\ feature (seen tentatively in the individual spectra for MIPS16080, MIPS22303 and MIPS22633) appears clearly present. The stacks show a possible 6.2\um\ feature as well; however, caution must be taken as for several of the sources this feature falls in the unreliable edge region. To test the reliability of this `detection' we also stacked only the sources with reliable spectra in this region (i.e. MIPS464, MIPS15880, MIPS16080, MIPS16152, MIPS22548). The 6.0\um\ absorption and the 6.85 and 7.25\um\ HAC features appear in this as well, although the 6.2\um\ `bump' is less pronounced than seen in Figure\,\ref{fig_stack}. 

The 4.67\um\ CO feature is hinted at but seems to be driven by the feature in MIPS464 as when this source is not included we have the four outlier stacks. This re-inforces the conclusion that, apart from the three sources discussed before, the CO feature is either not present or is extremely weak ($\tau_{4.6}$\,$<<$\,0.2) in the majority of our sample.  

\subsection{Silicate feature depth reliability \label{sec_si}}

The silicate feature does not fall within our new deeper data and is discussed in more detail in \citet{sajina07} for the GO1 sources and \citet{dasyra09} for the GO2 sources. However here we address the measurement and reliability of the silicate feature depths for our 11 targets specifically. This is necessary because in order to put our newly detected molecular absorption features in context with local sources we would like to look at the ice-to-silicate and HAC-to-silicate feature ratios (which we do in the Discussion).  Many of our sources have poorly defined continuum beyond rest-frame $\sim$\,11\um\ and also tend to have very low signal-to-noise inside the feature itself (see Figure\,\ref{fig_specs}). These features were originally fit with a power law plus extinction law model \citep[see][]{sajina07}, which as discussed earlier relates to the observed silicate feature depth via $\tau_{9.7,model}$\,=1.4\,$\tau_{9.7,obs}$.  For some of the sources this is fairly well determined, but for others the fitted value has large uncertainty (either due to poorly determined continuum, or uncertainty due to the PAH features flanking the feature). For example, for MIPS464 we derived model $\tau_{9.7}$\,=\,4.9\,$\pm$\,1.6 which translates to observed $\tau_{9.7,obs}$\,=\,3.4\,$\pm$\,1.1.  By contrast, MIPS15880 has considerably better defined feature, and we derive model $\tau_{9.7}$\,=\,3.9\,$\pm$\,0.4 which translates to observed $\tau_{9.7,obs}$\,=\,2.8\,$\pm$\,0.3. In the case of the strong-PAH source, MIPS22530, the silicate feature is saturated and we can only determine that it has observed silicate absorption of $\tau_{9.7}$\,$>$\,3.7. 

In order to further test the uncertainty associated with our silicate feature optical depth estimates, we also fit all of them using the same procedure as used for the local ULIRGs in \citet{spoon07}. We find that for the majority of the sources, the two approaches give consistent results, the values obtained differ most for the sources with poor continuum and noisy features, but are within the fitted uncertainty (e.g. MIPS464 has $\tau_{Si}$\,=\,2.5 consistent with our $\tau_{9.7}$\,=\,3.4\,$\pm$\,1.1). The most substantial difference comes for MIPS22530 where we obtain $\tau_{Si}$\,=\,1.7, considerably shallower than the value obtained from our model fitting. This is a well known uncertainty arising from the different treatment of the PAH features. Since MIPS22530 and MIPS22432 (the other strong-PAH source) do not have either  HAC or ice detections, they do not feature in our optical depth ratio estimates (see Discussion) and hence this substantial uncertainty does not affect our results. 

\section{Discussion}

\subsection{PAH properties at $z$\,$\sim$\,2}

\begin{figure}[h!]
\plotone{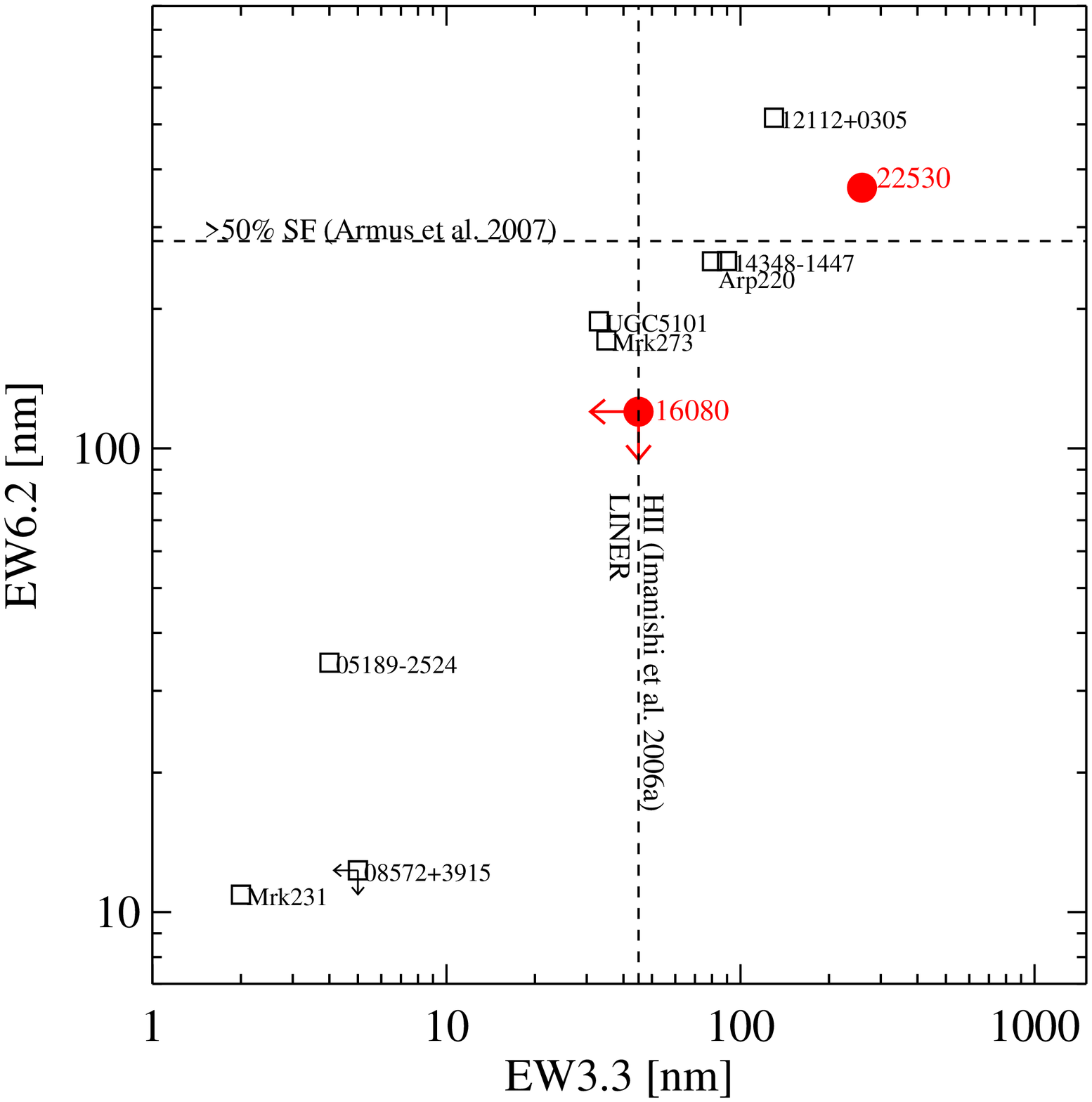}
\caption{The equivalent widths of the 3.3\um\ and 6.2\um\ PAH features in local ULIRGs compared with  MIPS22530 and MIPS16080 (red dots). The local ULIRG data as well as the criteria for starburst-dominated sources are based on \citet{imanishi_ulirgs} and \citet{armus07}. \label{pah_ratio}}  
\end{figure}

Figure~\ref{pah_ratio} shows the EW3.3 vs. EW6.2 for some representative local ULIRGs. We also show the derived values for MIPS22530 and MIPS16080 (which is an upper limit only as neither feature is unambiguously detected by itself).  We show the starburst-dominated source limits as determined by \citet{imanishi_ulirgs} based on optical spectral classification, and \citet{armus07} based on neon line ratios. On both counts,  MIPS22530 satisfies the starburst-dominated ULIRG criteria, consistent with our prior knowledge of this source. MIPS16080 on the other hand is more likely a composite source, as expected. 

Four of the sources showed signs of the 3.3\um\ PAH feature, with MIPS22530 having the most unambiguous detection. Based on the fits presented in \S\,4.6, we can estimate 3.3\um\ PAH log luminosities of 9.6, 9.6, 9.8 and 10.1 for MIPS16080, MIPS22303, MIPS22530, and MIPS22633.  The total 3\,--\,1000\um\ infrared luminosities for the first three sources are given in \citet{sajina08} and for MIPS22633 we have $\log(L_{\rm{IR}}/L_{\odot})$\,=\,12.9 (Sajina et al. in prep.). For our most reliable 3.3\um\ PAH source (MIPS22530), we have $L_{3.3}/L_{\rm{IR}}$\,=\,8.9\,$\times$\,$10^{-4}$. This is consistent with the $L_{3.3}$/$L_{\rm{IR}}$\,=\,8.5\,$\times$\,$10^{-4}$ reported by \citet{siana09} for a lensed $z$\,$\sim$\,3 Lyman Break galaxy, the only other high-$z$ measurement of the 3.3\um\ PAH. This ratio is somewhat larger than observed for local ULIRGs \citep[$\sim$\,$10^{-4}$;][]{imanishi_ulirgs}. However, this is believed to be an aperture effect, as this ratio is observed to increase with distance with the more distant ULIRGs having a ratio of $\sim$\,$10^{-3}$ \citep{mouri90}.   

For our other sources we find $L_{3.3}/L_{\rm{IR}}$\,=6.6\,$\times$\,$10^{-4}$ (MIPS16080), 4.3\,$\times$\,$10^{-4}$ (MIPS22303) and 1.6\,$\times$\,$10^{-3}$ (MIPS22633).  If we take a ratio of $\sim$\,$10^{-3}$ to be indicative of starburst-dominated systems \citep[see][]{siana09}, then the above can give us an alternative look into what fraction of the $L_{\rm{IR}}$ is due to starburst activity and also what SFRs are feasible. Accordingly, MIPS16080 likely has equal AGN and starburst contributions to the bolometric luminosity, whereas for MIPS22633, the starburst may well be dominant. In MIPS22303, the AGN is likely the dominant component. These are consistent with our earlier results \citep{sajina08}. Note however, that using the above translation between $L_{3.3}$ and $L_{\rm{IR}}$ implies starburst-only luminosities of $\gs$\,$10^{12}$\lsun\ even for MIPS16080 and MIPS22303, and consequently SFRs of $\sim$\,100\msun/yr.  However, as discussed in \S\,4.6, these features being of low signal-to-noise ratios and somewhat confused with the ice and HAC absorption features, are not particularly reliable. The stacked spectrum shows that PAH emission is likely present in the majority of these sources, even excluding the two strong-PAH sources. The fact that star-formation is present is further supported by the fact that although most of these sources are not individually detected in the far-IR,  their stacked 1.2mm (rest-frame $\sim$\,400\um) flux is strongly detected at $\sim$\,0.5\,mJy which is comparable to fainter ($S_{850}$\,$\sim$\,2\,--\,3\,mJy) sub-mm galaxies \citep{sajina08}.  These numbers suggests that SFRs on the order of 100\,\msun/yr are feasible for the bulk of the weak-PAH sample.  Although in itself significant, star-formation at this level is insufficient to dominate the bolometric luminosities of these sources. 

\subsection{Ice and HAC detectability compared with local ULIRGs}

Local ULIRGs show a high level of detectability of water ice absorption features. For example the 46 sources in the AKARI sample \citep{imanishi08} were selected out of the IRAS 1\,Jy sample merely on the basis of visibility to AKARI and are therefore unbiased with respect to the local IRAS ULIRG population as a whole. The 3\um\ feature was detected in 30/46 sources (65\%), whereas the 3.4\um\ HAC feature was only observed in 4 of the sources (9\%).  The presence of the HAC feature has been suggests as an indicator of AGN-dominance in the mid-IR \citep{risaliti06,imanishi_ulirgs} and hence this low fraction is consistent with local ULIRGs being predominantly starbursts. To determine the appropriate detectability for our sample we need to consider first that MIPS8392 has too low a signal to noise in this regime, and three other sources are of $z$\,$\ls$\,1.8 which makes  this feature too near the edge for reliable detections.  This means that 3/7 (43\%) of our sample show  the 3\um\ water ice absorption feature, this is roughly comparable to the local ULIRGs given the small number of sources in our sample. On the other hand (5/11) of the sources (45\%) show the 3.4\um\ HAC absorption feature, which fraction is significantly higher than observed for local ULIRGs (even accounting for the Poisson error in such a small sample). However, the 3.4\um\ HAC feature can be practically difficult to detect in sources with strong 3.0\um\ ice and/or 3.3\um\ PAH emission. This is one of the explanation for the low HAC detectability among the local ULIRGs \citep{imanishi08}. 

We have already shown that the best analogs to our sources are the sources with red 3\,--\,4\um\ spectra rather than the more typical local ULIRGs. There are five sources in the \citet{imanishi08} AKARI sample that meet our criteria (i.e. $\alpha_{3-4}$\,$>$\,2). Of these, 3 show the HAC feature, and a different set of 3 show the 3.0\um\ ice feature.  This is roughly in better agreement with the observed detection rates for our sample (but with very poor statistics!).

\begin{figure}[h!]
\plotone{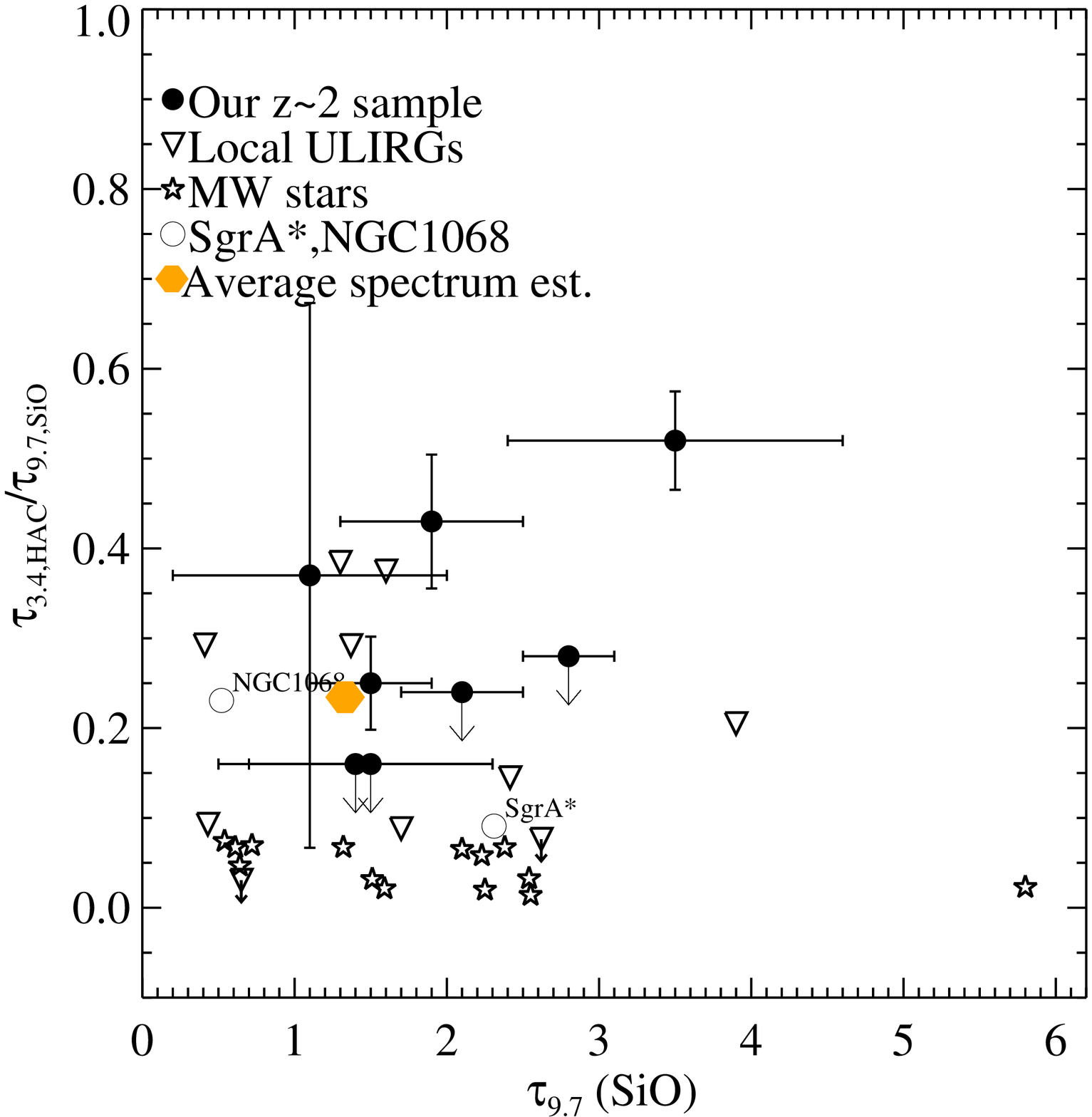}
\caption{The ratio of the 3.4\um\ HAC optical depth to 9.7\um\ silicate feature optical depth vs. the 9.7\um\ silicate feature.  For comparison we show the data for embedded protostars \citep{gibb04} and local ULIRGs (see text). We also indicate the location of SgrA* \citep{gibb04} and NGC1068 \citep{imanishi00}. The orange hexagon shows the average value for the sample based on the stacked spectra (but see \S\,4.8 for caution). \label{hac_si}}  
\end{figure}

\subsection{HAC-to-silicate ratio}

Figure\,\ref{hac_si} shows the HAC/silicate ratio vs. the silicate feature strength for our sources compared with local ULIRGs. To construct this figure, we use the $\tau_{\rm{3.4}}$ measurements for local ULIRGs from several studies \citep{imanishi_ulirgs,risaliti06,sani08,imanishi08} combined with the silicate feature depths from \citet{armus07} and \citet{sirocky08}. We find that, while the HAC detectability of our sample is higher than the local ULIRG population, the HAC/silicate ratio appears consistent with that observed among local ULIRGs where the HAC feature is detected. 

 For comparison we also show the data for embedded protostars from the review paper of \citet{gibb04}. Note that the HAC-to-silicate ratio observed in the MW is remarkably constant across a wide range of silicate feature depth. This is consistent with the diffuse ISM value of 0.05\,--\,0.06 \citep{pendleton94}. Moreover, a recent study of the spatial variation of the optical depth in the Circinus galaxy (the nearest AGN to the MW) found that although the absolute optical depth changed dramatically, this ratio remained nearly constant \citep{colling09}. An enhanced HAC-to-silicate ratio was first reported by \citet{imanishi00} who studied a sample of 4 local obscured AGN nuclei including NGC1068 (shown in Figure\,\ref{hac_si}). Their favored interpretation of this enhanced ratio comes from the following explanation: each of the two optical depth samples the column density between us and the sources. If the dust is well mixed and in a screen approximation is valid the ratio is expected to remain constant (reflecting the relative abundance of the silicate and carbon grains and their optical properties). However if a steep temperature gradient between the $\sim$\,300\,K dust at 9.7\um\ and $\sim$\,1000\,K dust at 3.4\um\ is present, since the hotter dust is spatially closer to the power source, the HAC feature arises from a larger column density than the silicate feature and hence the enhanced ratio.  This is what is commonly referred to as a `buried nucleus' \citep[see Fig.2][]{imanishi_irs}. Notice that this model addresses only the obscuration geometry: it is neither a statement about the nature of the power source (although commonly taken to be an AGN in these cases), nor about the absolute level of obscuration (the embedded protostars show comparable silicate feature depths). 

An alternative explanation might be that deeply obscured ULIRGs and our $z$\,$\sim$\,2 sources simply have a higher proportion of hydrocarbons. For example, bombardment with H atoms can hydrogenate amorphous carbon grains increasing the HAC-to-silicate ratio \citep[see e.g.][]{gd00,duley05}. Various other processes can be important, s.a. UV radiation processing of ice mantles can lead to refractory materials on the grain surfaces that lead to the 3.4\um\ absorption feature \citep{greenberg95}, consistent with the somewhat weaker ice features. Ultimately, the degree of HAC absorption depends on the balance of processes that support hydrocarbon generation to those that suppress it (s.a. converting hydrocarbons to aromatics). It is not unreasonable to suppose that such processes might be more efficient in our sources than the typical Milky Way sources. The geometric explanation is simpler and therefore favored, however, a difference in composition cannot be excluded. 

\begin{figure}[h!]
\plotone{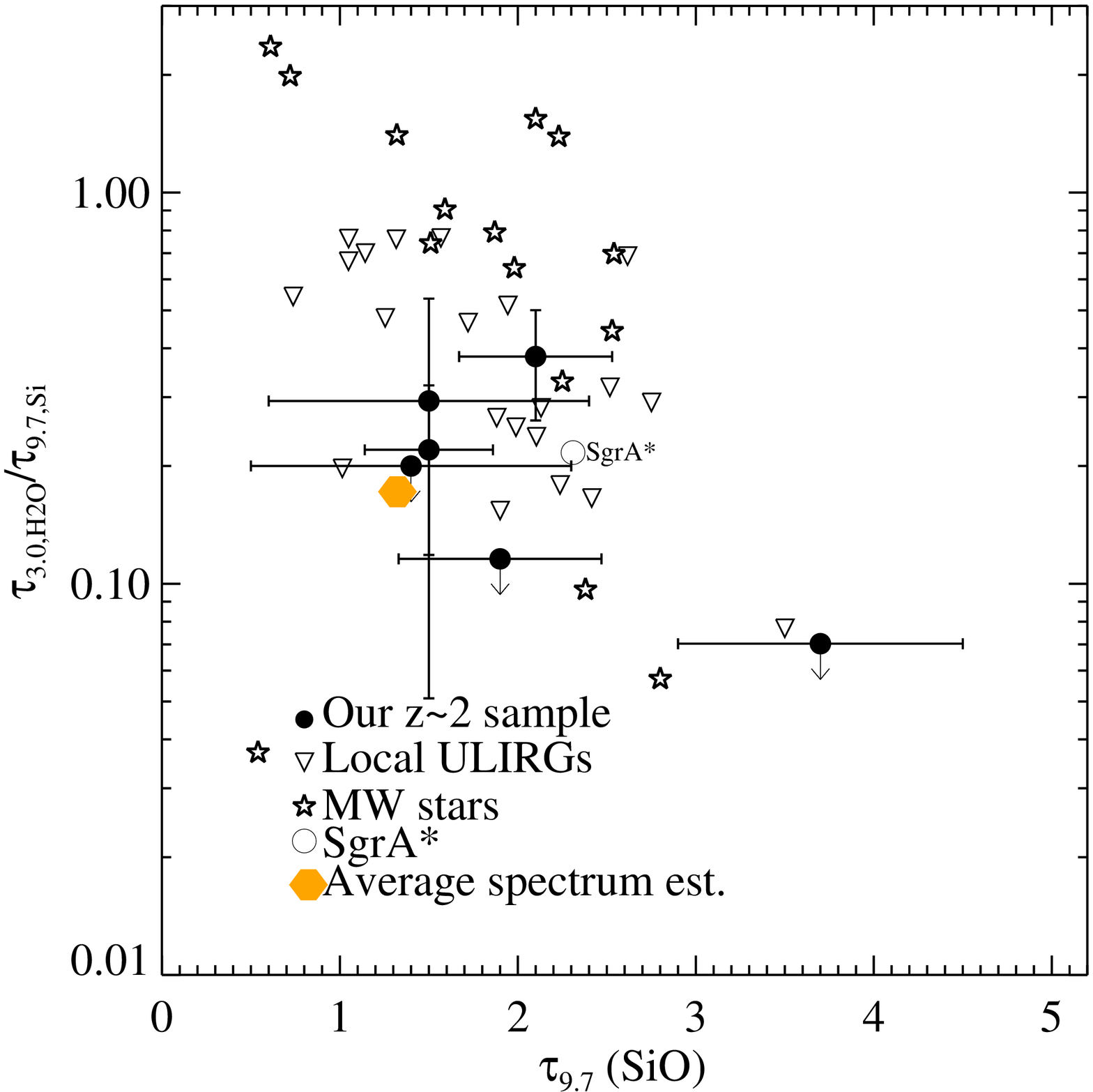}
\caption{ The ratio of the water ice optical depth to silicate feature depth. Symbols are as in Figure\,\ref{hac_si}. The orange hexagon shows the average value for the sample based on the stacked spectra (see also \S\,4.8). The dashed lines represent the approximate ratios expected for different ice mantle thicknesses (as indicated). Even when ice absorption is detected, our sources are consistent with fairly thin ice mantles.  \label{ice_si}}  
\end{figure}

\subsection{Ice to silicate ratio}
Figure\,\ref{ice_si} shows the ratio of the 3.0\um\ ice feature optical depth to the 9.7\um\ silicate dust optical depth. As before, we compare our sample with both local ULIRGs and the data for MW stars. We find that unlike the case for the HAC-to-silicate ratio which is remarkably similar across for all available MW sources, the ice-to-silicate ratio shows a wide spread of more than an order of magnitude. This ratio is strongly dependent on the ice mantle thickness \citep[e.g.][]{oss94,zinoveva05}, but radiative transfer details following the arguments presented in the previous section likely dominate. 
When we consider only the sources with red 3\,--\,4\um\ spectra, and where both ice and silicate optical depths are available, we find that IRAS08572+3915 has $\tau_{3.0}/\tau_{9.7}$\,=\,0.08, and IRAS19254-7245 has   $\tau_{3.0}/\tau_{9.7}$\,=\,0.15. Although a much larger sampling is needed, this suggests (especially in conjunction with our own sample) that sources with redder 3\,--\,4\um\ slopes have lower observed ice\,-\,to\,-\,silicate ratios. 
 
\subsection{The 4.67\um\ CO feature \label{sec_co2}}
The presence of CO gas in at least some of these systems is known independently through mm-wave spectroscopic observations (Yan et al., in prep.).  In particular, MIPS16080 and MIPS22530 both show significant masses of cold CO gas. Unfortunately both of these are among the sources, where the 4.67\um\ feature falls in the unreliable parts of the co-added spectra. Moreover, the presence of CO does not automatically imply the presence of the 4.67\um\ CO feature which arises from the much warmer regions in the vicinity of the galactic nuclei.  

We would expect to see this feature in our sources as it is seen in some deeply obscured local ULIRGs \citep{spoon_co}. We do show some evidence for the 4.67\um\ CO absorption feature in three of these sources.  Although tentative, this is the first suggestion of this feature in such high redshift systems. However this feature is very weak even in these sources and only seen in 3 of the 7 sources with usable spectra (and is not present in the stacked spectra).  We can therefore conclude that it is either not ubiquitous even among sources with silicate absorption, or more realistically that the feature manifests with a range of absorption depths, and we only see the most pronounced features. 

As discussed in \S\,4.5, a $\sim$\,5\um\ starburst continuum would effectively weaken the observed CO feature depth (as that likely originates close to the AGN itself).  For weak-PAH sources, the $\sim$\,5\um\ continuum is most likely dominated by the AGN  \citep{sajina07}, however a starburst contribution, sufficient to affect the CO depth, cannot be excluded.  The fact that one of our two strong-PAH sources (MIPS22432) is also one of three sources showing some hints of CO absorption, suggests that this is much more complicated that a simple correction scaled to the starburst strength (as revealed by the PAH).   

Drawing a parallel with the dust-embedded Galactic YSOs, perhaps there is a range from purely solid form CO to fully gaseous CO as the temperatures rise and the CO previously deposited on grain surfaces sublimates \citep[see][; and references therein]{whittet03}. Solid-phase CO shows a much narrower feature than its gas-phase counterpart, and in our low-resolution spectra would be undetectable, unless exceptionally strong. In local ULIRGs, this problem is present as well, and only the high resolution spectrum of IRAS08572+3915 has allowed to rule out CO ice in this source \citep{geballe06}. To date, only NGC4945 has a CO ice detection \citep{spoon03}. 

Following the argument in \citet{lutz04}, we might also conclude that we are not seeing this feature due to dust tori obscuration rather than `buried nuclei'. However, this runs counter to our conclusions from the strong 3.4\um\ HAC absorption features (see \S\,5.3).  Ultimately, there are too few 4.67\um\ CO detections even locally (see Introduction) and therefore the statistics on what fraction of the sources we should expect to show this feature sufficiently strongly are lacking. 

\subsection{Buried nuclei at $z$\,$\sim$\,2}

Our sources are drawn from sample of 24\um\ flux limited and color-selected sources \citep{yan07}. The 0.9\,mJy 24\um\ flux limit already restricts us at $z$\,$\sim$\,2 to the highest luminosity sources, and indeed we find $\langle L_{\rm{IR}}\rangle$\,$\sim$\,$10^{12.7}$\lsun\ for all our $z$\,$\sim$\,2 sources \citep{sajina08}.  \citet{imanishi09} shows that locally, at comparable luminosities, $\sim$\,70\% of the ULIRGs host buried AGN, compared to $\sim$\,30\% in lower luminosity ULIRGs. It is not surprising, therefore, that in \citet{sajina07} we found that the bulk of our $z$\,$\sim$\,2 sources showed AGN-like mid-IR spectra, usually accompanied by silicate absorption features. Because of the biased selection of our sample, we cannot test whether or not obscured AGN form a larger fraction of $z$\,$\sim$\,2 ULIRGs than observed for the local ULIRGs.  We did show that, consistent with the well-known luminosity evolution between $z$\,$\sim$\,2 and today, the number density of our silicate-absorption sources is about an order of magnitude higher than observed locally (for ULIRGs with comparable silicate absorption \citep{sajina07}). However, silicate absorption alone is not a strong enough diagnostic of buried AGN since it can in principle arise from a range of different spatial scales (i.e. host obscuration).

In this study, we find that beyond the presence of silicate absorption, the full 2-8\um\ SEDs of our $z$\,$\sim$\,2 sources, the presence and strength of 3.4\um\ amorphous hydrocarbons and 3.0\um\ water ice absorption features are also comparable to those observed in nearby heavily obscured ULIRGs. This suggests similar obscuration geometries. Their closest counterparts in the local  Universe are believed to be dominated by buried nuclei (i.e. compact sources with steep dust temperature gradients) most likely harboring AGN as discussed in the preceeding sections \citep[see e.g.][for a more complete discussion]{imanishi_irs}. The lack of strong 4.67\um\ CO absorption however is inconsistent with this picture, although the dearth of such observations locally makes it difficult to interpret (see \S\,5.5).  We find evidence of PAH emission even among the weak-PAH (i.e. AGN-dominated in the mid-IR)  sources, suggesting that star-formation is present at least in some of these sources. Scaling from the 3.3\um\ luminosities we expect that SFRs of order $\sim$\,100\msun/yr could be present. This is typically insufficient to dominate the bolometric output of these galaxies. The combination of deeply obscured, likely quasar, nuclei and ULIRG-strength star-formation supports the view of co-evolution of black-hole growth and host galaxy stellar mass growth suggested by a variety of observational and theoretical considerations \citep[see e.g.][]{magorrian98,cs01,phop08,ams08}. 

\section{Summary \& Conclusions}

1) In this paper we present deep rest-frame 2\,--\,8\um\ spectra for 11 $z$\,$\sim$\,2 ULIRGs. \\

2) Our sources tend to have redder 3\,--\,4\um\ continua than local ULIRGs. The few local sources with comparably red spectra, show continuum-dominated 2-10\um\ spectra with weak-PAH and strong silicate absorption (consistent with their being predominantly AGN-dominated). They are therefore overall good analogs to our sources. \\

3) We found that nearly half of the sources with sufficient coverage (43\%) show water ice absorption at 3\um\ (compared with 65\% for local ULIRGs). There is no clear trend of detectability with increasing silicate feature depth. \\ 

4) The 3.4\um\ HAC absorption is seen in 40\% of the sample compared with 9\% for local ULIRGs. The HAC-to-siilcate ratio, $\tau_{3.4}/\tau_{9.7}$, is 2\,--\,5 times stronger that the typical ratio observed in the Milky Way, but consistent with the obscured nuclei of some local ULIRGs. This is consistent with compact nuclei buried in dust with a steep temperature gradient distribution.   \\

5) Three of our sources show tentative signs of the CO 4.67\um\ feature. If genuine CO absorption, these features show a strong asymmetry between the short- and long-$\lambda$ branches of this feature. Moreover, the observed strength is still significantly lower than observed in local ULIRGs. Stacking analysis suggests this feature is not present in the bulk of our sample or is very weak ($\tau_{4.6}$\,$<<$\,0.2). \\
 
6) One source, MIPS22530, has a clear 3.3\um\ PAH emission feature detected. The ratio of the 6.2 to 3.3\um\ PAH strengths is consistent with that observed in local ULIRGs, and the overall strength of both features implies a starburst-dominated source, consistent with earlier results. Three more sources (MIPS16080 MIPS22303, and MIPS22633) show tentative evidence for the 3.3\um\ PAH.  The stacked spectra show evidence for PAH emission in the bulk of the weak-PAH sources implying likely significant star-formation in conjunction with the obscured black-hole accretion. \\

\acknowledgements
We thank the anonymous referee for their thoughtful comments, which have improved this paper. We also wish to thank Brian Siana and Kalliopi Dasyra for helpful discussions. 
This work is based on observations made with the {\sl Spitzer} Space Telescope, which is operated by the Jet Propulsion Laboratory, California Institute of Technology under a contract with NASA. Support for this work was provided by NASA through an award issued by JPL/Caltech.

\clearpage

\begin{deluxetable}{cccccccc}
\tablecolumns{8}
\tablewidth{6.2in}
\centering
\tabletypesize{\scriptsize}
\tablecaption{\label{table_obs} Summary of observations }
\tablehead{\colhead{MIPSID} &  \colhead{$z_{old}\tablenotemark{a}$}  &\multicolumn{2}{c}{G01}  & \multicolumn{2}{c}{G02} & \multicolumn{2}{c}{G04} \\
\colhead{} & \colhead{} & \colhead{SL1} & \colhead{LL2} &   \colhead{SL1} & \colhead{LL2} & \colhead{SL1} & \colhead{LL2} }
\startdata
464 & 1.85 & -- & 6\,$\times$\,120\,s (S)\tablenotemark{b} & --- & --- & 4\,$\times$\,40\,$\times$\,60\,s (M) & 30\,$\times$\,120\,s (S) \\
8392\tablenotemark{c} & 1.9 & 2\,$\times$\,60\,s(S) & --- & --- & 12\,$\times$\,120\,s (S) & -- & 30\,$\times$\,120\,s (S) \\
15880 & 1.68 & --- & 6\,$\times$\,120\,s (S) &  --- & --- & 4\,$\times$\,8\,$\times$\,60\,s (M) & -- \\
16059 & 2.325 & --- & 6\,$\times$\,120\,s (S) &  --- & --- & 4\,$\times$\,40\,$\times$\,60\,s (M) & 30\,$\times$\,120\,s (S) \\
16080 & 2.007 & --- & 6\,$\times$\,120\,s (S) &  --- & --- & 4\,$\times$\,40\,$\times$\,60\,s (M) & 30\,$\times$\,120\,s (S) \\
16152 & 1.83 & --- & --- & --- & 6\,$\times$\,120\,s (S) & 6\,$\times$\,60\,s (S) & 30\,$\times$\,120\,s (S) \\
22303 & 2.34 & --- & 6\,$\times$\,120\,s (S) & --- & --- & 4\,$\times$\,8\,$\times$\,60\,s (M) & 30\,$\times$\,120\,s (S) \\
22432 & 1.59 & --- & --- & 1\,$\times$\,240\,s(S) & 6\,$\times$\,120\,s(S) & -- & --- \\
22530 & 1.952 & --- & 6\,$\times$\,120\,s (S) & --- & --- & 4\,$\times$\,40\,$\times$\,60\,s (M) & 30\,$\times$\,120\,s (S) \\
22548 & 2.12 & --- & --- & --- & 6\,$\times$\,120\,s (S) & 6\,$\times$\,60\,s (S) & 30\,$\times$\,120\,s (S) \\
22633 & 2.36 & --- & --- & --- & 6\,$\times$\,120\,s (S) & 4\,$\times$\,8\,$\times$\,60\,s (M) &  ~30\,$\times$\,120\,s (S) 
\enddata
\tablenotetext{a}{Based on \citet{sajina07} and \citet{dasyra09}.}
\tablenotetext{b}{We use the convention (\#positions\,$\times$\,)\#cycles\,$\times$\,ramp duration. In parentheses we mark whether the data is taken in stare mode (S) or spectral mapping (M). For simplicity the number of positions (i.e. the 2 nod positions) is omitted for stare mode observations. }
\tablenotetext{c}{This source is identified as FLS CVZ VLA9 from a GO1 GTO program (PID\#15, PI J. Houck). It is also identical to AMS06 \citep{ams08}.}
\end{deluxetable}

\begin{deluxetable}{ccccc}
\tablecolumns{5}
\tablewidth{3.5in}
\tabletypesize{\scriptsize}
\tablecaption{\label{table_slopes} Continuum parameters}
\tablehead{\colhead{MIPSID} & \colhead{$L_{3\mu m}$} & \colhead{$L_{5.5\mu m}$} & \colhead{3\,--\,4\um\ slope}  & \colhead{5\,--\,7\um\ slope}}
\startdata
    464 & 11.07 & 11.66 &  3.12 &  2.85 \\
    8392 & 11.63 & 12.11 & 3.63 & 0.91 \\ 
   15880 & 11.31 & 11.89 &  2.52 &  2.90 \\
   16059 & 11.74 & 12.12 &  3.15 &  0.97 \\
   16080 & 11.49 & 11.95 &  3.38 &  0.95 \\
   16152 & 11.68 & 11.90 &  2.12 &  1.59 \\
   22303 & 11.95 & 12.42 &  3.30 &  1.08 \\
   22432 & 11.41 & 11.90 &  3.75 &  2.76 \\
   22530 & 11.10 & 11.51 &  1.90 &  4.76 \\
   22548 & 11.88 & 12.03 &  2.26 &  1.54 \\
   22633 & 11.54 & 12.06 &  3.37 & -0.30 
\enddata
\end{deluxetable}

\begin{deluxetable}{cccccc}
\tablecolumns{6}
\tablewidth{4in}
\tabletypesize{\scriptsize}
\tablecaption{\label{table_taus} Optical depth of absorption features}
\tablehead{\colhead{MIPSID} & \colhead{$z_{\rm{old}}$} & \colhead{$z_{\rm{new}}$} & \colhead{$\tau_{3.05}$} & \colhead{$\tau_{3.4}$} & \colhead{$\tau_{9.7}$} }
\startdata
464 & 1.85 & 1.75 & --- & 0.52\,$\pm$\,0.10\tablenotemark{a}  & 3.5\,$\pm$\,1.1\tablenotemark{b} \\
8392 & 1.90 & 1.90 & --- & $<$\,0.45 & $>$\,4.9 \\
15880 & 1.68 & 1.62 & --- & $<$\,0.28 & 2.8\,$\pm$\,0.3 \\
16059 & 2.325(Keck) & 2.325 & $<$\,0.22 & 0.43\,$\pm$\,0.04 & 1.9\,$\pm$\,0.6 \\
16080 & 2.007(Keck) & 2.007 & 0.33\,$\pm$\,0.13 & 0.25\,$\pm$\,0.04 & 1.5\,$\pm$\,0.4 \\
16152 & 1.83 & 1.84 & $<$\,0.28 & $<$\,0.16 & 1.4\,$\pm$\,0.9 \\
22303 & 2.34 & 2.28 & 0.80\,$\pm$\,0.19 & $<$\,0.24 & 2.1\,$\pm$\,0.4 \\
22432 & 1.59 & 1.60 & --- & --- & 1.8\,$\pm$\,0.6 \\
22530 & 1.952(Keck)  & 1.952 & $<$\,0.34 & --- & $>$\,3.7 \\
22548 & $>$\,2.12 & 2.15 & $<$\,0.26 & $<$\,0.16 & 1.5\,$\pm$\,0.8 \\
22633 & 2.36 & 2.20 & 0.44\,$\pm$\,0.25 & 0.37\,$\pm$\,0.14 & 1.1\,$\pm$\,0.9 
\enddata
\tablenotetext{a}{The errors on the ice and HAC features only reflect the per pixel errors in the spectra, assuming that errors in the continuum levels are negligible. We believe this is a reasonable assumption, in all cases except MIPS22633, as can be seen in  Figure\,\ref{buried_conts}. }
\tablenotetext{b}{The errors on $\tau_{Si}$ are based on the MCMC fitting as described in \citet{sajina07}. They therefore include the per pixel errors, the uncertainty in the PAH+continuum decomposition, and the frequently poorly determined continuum above $\sim$\,11\um.  For strong-PAH sources however, the depth of this feature is less well defined than shown here (see \S\,\ref{sec_si} for further details).}
\end{deluxetable}

\end{document}